\documentclass[a4paper, 12pt, twocolumn]{article}

\usepackage[top=83pt,bottom=83pt,left=47pt,right=47pt]{geometry}
\usepackage{url}

\usepackage{flushend}
\usepackage{layout}
\usepackage[dvipsnames]{xcolor}
\usepackage{lipsum}
\usepackage{graphicx}

\usepackage[small]{caption}
\usepackage{subcaption}
\usepackage{amssymb}
\usepackage{mathtools}
\usepackage{xspace}
\usepackage{upgreek}
\usepackage{setspace}

\usepackage{fancyhdr}

\usepackage[switch]{lineno}
\fancypagestyle{firststyle}
{
   \fancyhf{}
   \fancyhead[LO,RE]{\textsc{Preprint} (Typeset: \date{\today}) \\
   \textbf{Direct detection of dark matter: a critical review}}
   \fancyfoot[LO,RE]{{\bf Authors' e-mails: }  \texttt{marcin.misiaszek@uj.edu.pl, nicola.rossi@lngs.infn.it}}
}

\usepackage{hyperref}
\hypersetup{colorlinks,bookmarksopen,bookmarksnumbered,
linkcolor=BlueViolet,pdfstartview=FitH,urlcolor=black,citecolor=BlueViolet}

\providecommand{\keywords}[1]
{
  \small    
  \textbf{\textit{Keywords---}} #1
}

\title{\color{BlueViolet} \bf DIRECT DETECTION OF DARK MATTER: \\ A CRITICAL REVIEW}
\author{Marcin Misiaszek\textsuperscript{a} and Nicola Rossi\textsuperscript{b} \\
\textsuperscript{a}{\footnotesize Jagiellonian University, Krakow, Poland} \\
\textsuperscript{b}{\footnotesize Laboratori Nazionali del Gran Sasso (INFN), L'Aquila, Italy}
}
\date{\empty}

\begin{document}

\pagestyle{fancy}
%... then configure it.
\fancyhead{} % clear all header fields
\fancyhead[LO,RE]{M. Misiaszek \& N. Rossi}
\fancyhead[RO,LE]{DIRECT DETECTION OF DARK MATTER}
\fancyfoot{} % clear all footer fields
\fancyfoot[LE,RO]{\thepage}
\fancyfoot[LO,RE]{}

\twocolumn[
\begin{@twocolumnfalse}

\maketitle

\thispagestyle{firststyle}

\begin{abstract}
The nature of the dark matter in the Universe is one of the hardest unsolved problems in modern physics. Indeed, on one hand, the overwhelming indirect evidence from astrophysics seems to leave no doubt about its existence; on the other hand, direct search experiments, especially those conducted with low background detectors in underground laboratories all over the world seem to deliver only null results, with a few debated exceptions. Furthermore, the lack of predicted candidates at the LHC energy scale has made this dichotomy even more puzzling. We will recall the most important phases of this novel branch of experimental astro-particle physics, analyzing the interconnections among the main projects involved in this challenging quest, and we will draw conclusions slightly different from how the problem is commonly understood.  
\end{abstract}


\vspace{1.5cm}
\end{@twocolumnfalse}
]

\section*{Introduction}

According to Standard Cosmology, the visible Universe emerged, after a very short phase of rapid inflation, from a space-time singularity, with infinite temperature and density, about 13.8 billion years ago, and its evolution from that moment is determined by the distribution of its energy content among its distinguished constituents, namely visible (or baryonic) matter (4.9\%), dark (non visible) matter (26.5\%) and dark (non visible) energy (68.5\%)~\cite{bib:PDG}.

Besides dark energy, whose physical nature is still a subject of intense debate, dark matter offers more specific clues. These clues have led the scientific community to strongly believe in its existence and even to consider possible ways to experimentally detect its presence when it passes through the Earth.

The dark matter hypothesis fills the gap in the explanation of many astrophysical observations, from gravitational collapse of structures in the early Universe, patterns in cosmic microwave background, anomalous dynamics and gravitational lensing of astronomical objects from small-sized galaxies to large super-clusters, with some exceptional case as the Bullet Cluster, considered by someone as an incontrovertible smoking gun.
These strong experimental evidences seem to converge towards the existence of a given amount of invisible massive particles piled up around massive astronomical objects with some rudimentary properties.

In the dominant paradigm, this matter, indeed,  should have had a negligible velocity at the time of early structure formation, for this reason is called \emph{cold}, as stated by the widely accepted \emph{Lambda Cold Dark Matter} paradigm ($\Uplambda$CDM)~\cite{bib:LCDM}. This scenario can be easily actualized with the existence of heavy and collision-less particles, \emph{e.g.} massive particles with mass comparable or larger than typical atomic nuclei (tens of GeV/c$^2$ or more). If massless or much lighter, the dark matter should have some interaction properties, or some coherent aggregation, sufficient to simulate a non-relativistic gravitational clustering, capable of speeding up the structure formation, as observed, after the matter-radiation decoupling epoch in the early Universe~\cite{bib:pad}.

Alternative theories, as the \emph{Modified Newtonian Dynamics} (MOND)~\cite{bib:MOND}, super-fluid dark matter~\cite{bib:sfluid}, or even full General Relativity approach~\cite{bib:GRgaia}, have been showing waves of interest in the scientific literature. It is worth mentioning, even though it is not the main topic of this article, that some empirical evidences, as the Tully-Fisher relation~\cite{bib:tully} and the Renzo's rule~\cite{bib:renzo} for galaxies are hardly explainable by standard cold dark matter galactic halos, and this issues, very well addressed by the MOND theories,  seem sometimes completely ignored when considering whatever particle explanation of the missing mass of the Universe. And, along with this, many issues related to the $\Uplambda$CDM model are still unsolved, such as the \emph{missing satellites} problem, the \emph{cusp-core} problem and \emph{too-big-to-fail} problem (see \emph{e.g.}~\cite{bib:PDG} and Refs. therein).

Nevertheless, the particle hypothesis for dark matter, even if strongly weakened by the astonishing absence of the super-symmetry (SUSY) at the Higgs scale in the LHC accelerator \cite{bib:SUSY}, shows still a constant interest in the scientific community, mostly for its experimental feasibility. In other words, the costs and technology for the direct dark matter search in underground laboratories is still so affordable that the biggest part of the experimental community believes that it is still worth trying, as the main goal in scientific programs for the forthcoming decades.

Since the early 90s, when the hypothesis of dark matter began to be taken seriously by scientists, many candidates have been proposed, with mass ranging from axions with masses starting from $10^{-22}$ eV/c$^2$ to primordial black holes with masses up to 5 M$_{\odot}$ $\approx$ ($10^{67}$ eV/c$^2$)~\cite{bib:PDG}. This wide range of 89 orders of magnitude is filled almost homogeneously with many variants of the proposed models, and it is basically limited by astrophysical constraints related to the observed macro-structures. Within this range, it is worth mentioning the most important candidates with increasing mass: axions, with sub-eV/c$^2$ mass, detectable in haloscopes or low background detectors~\cite{bib:PDG};
sterile neutrinos~\cite{bib:sterili}, with mass of the order 1 keV/c$^2$, evergreen candidates for many presumed anomalies, never detected; weakly interacting massive particles (WIMP)~\cite{bib:wimp1, bib:wimp2}, with mass in the range 1--$10^3$ GeV/c$^2$, predicted by the SUSY extension of particle Standard Model (SM) and considered the top reference model almost up to the first null results by LHC~\cite{bib:lhc}, with some monster extensions up to the Grand Unification scale (GUT) of $10^{15}$ GeV/c$^2$ often called WIMPzillas~\cite{bib:zilla}, yet never detected; finally, dark objects (MACHOs) with the size of a planet ($\sim 10^{24}$  kg) or so~\cite{bib:machos}, and primordial black holes, with huge mass, up to 5 $M_\odot$ ~\cite{bib:BHlim}.
Finally, it is also worth mentioning the Mirror Matter model~\cite{bib:zurab}, conceptually different from WIMPs, but predicting the existence of dark matter candidates with mass comparable with the mass of visible chemical elements (1--100 GeV/c$^2$).

Another problem is the cross section scale between visible and dark matter. If one assumes that the typical cross section of the weak interaction ($\sigma\sim10^{-44}$ cm$^2$, or so), is already experimentally at reach, the range of possible interaction cross sections is actually really large. If one takes the squared Planck length as lower bound $\sigma\sim10^{-66}$ cm$^2$, there are 20 possible orders of magnitude, even though not all of them are actually testable.

All the models discussed above,  basically agree on the \emph{invisible} (electrically neutral) nature of the dark matter candidate: astrophysical observations constrain its possible electric charge and self-interaction, and require temporal stability with a lifetime comparable to the age of the Universe~\cite{bib:PDG}.

Indirect dark matter searches, such as hidden channels in accelerators or annihilation/decay in visible diffused particle backgrounds in the Galaxy, have also yielded no results~\cite{bib:PDG}.

\begin{table}[ht!]
    \centering
    \begin{tabular}{|c|c|}
  	   \hline
  	   \textsc{Property} & \textsc{Value}\\
  	   \hline
  	   Composition & -- \\
  	   Statistics & -- \\
  	   Family & -- \\
  	   Generation & -- \\
  	   Interaction & -- \\
  	   Symbol & $\chi$ \\
  	   Antiparticle & -- \\
  	   Mass & $10^{-22}$ eV $\div 5$ M$_{\odot}$ \\
  	   Half-life & $\gtrsim 10^{10}$ y \\
  	   Electric charge & $\lesssim  10^{-7}$ e [at 1 GeV] \\
  	   Self interaction & $< 0.5$ cm$^2$ [at 1 GeV] \\
  	   Magnetic Moment & -- \\
  	   Spin & -- \\
  	   Weak isospin & LH: --, RH:-- \\
  	   Weak hypercharge & LH: --, RH:-- \\
  	   Others & -- \\  	 
  	   \hline
    \end{tabular}
    \caption{Properties of the dark matter particle candidate to date. The choice of the symbol $\chi$ is not universal, but it is assumed in the present article as a reference for a generic dark matter candidate. From the top: composition, Bose or Fermi statistics, particle generation or family, fundamental interaction, symbol, antiparticle, mass, half-life, electric charge, self interaction, magnetic moment, spin, weak isospin, weak hypercharge and other characteristics. Symbol ``--'' stands for \emph{not available}.}
    \label{tab:dm}
\end{table}

Table~\ref{tab:dm} summarizes the known physical properties of dark matter particle candidates. The Table is essentially empty, except for some naive properties inferred by indirect astrophysical observations.
The absence of experimental evidence of dark matter in either direct or indirect search, has sometimes induced the literature to hide or rename some of the historical candidates, often with fancy names as ``axion-like" (ALPs) or ``WIMP-like"~\cite{bib:alps}, just because the original proposed theory seemed not to hold any more in upgraded experimental contexts. Furthermore, often there are unconscious biases, like the belief that, if the dark matter is not found in the range of the WIMPs, maybe it is really important to search for some \emph{light} WIMP-like dark matter particle with mass below 1 GeV/c$^2$, down to a few tens of MeV/c$^2$. This kind of suggestion is of course much weaker, if one considers that the theoretical prior on some plausible model, from $10^{-22}$ eV/c$^2$ to $5$ M$_\odot$ is currently basically uniform.
\begin{figure}
    \centering
    \includegraphics[width=\columnwidth]{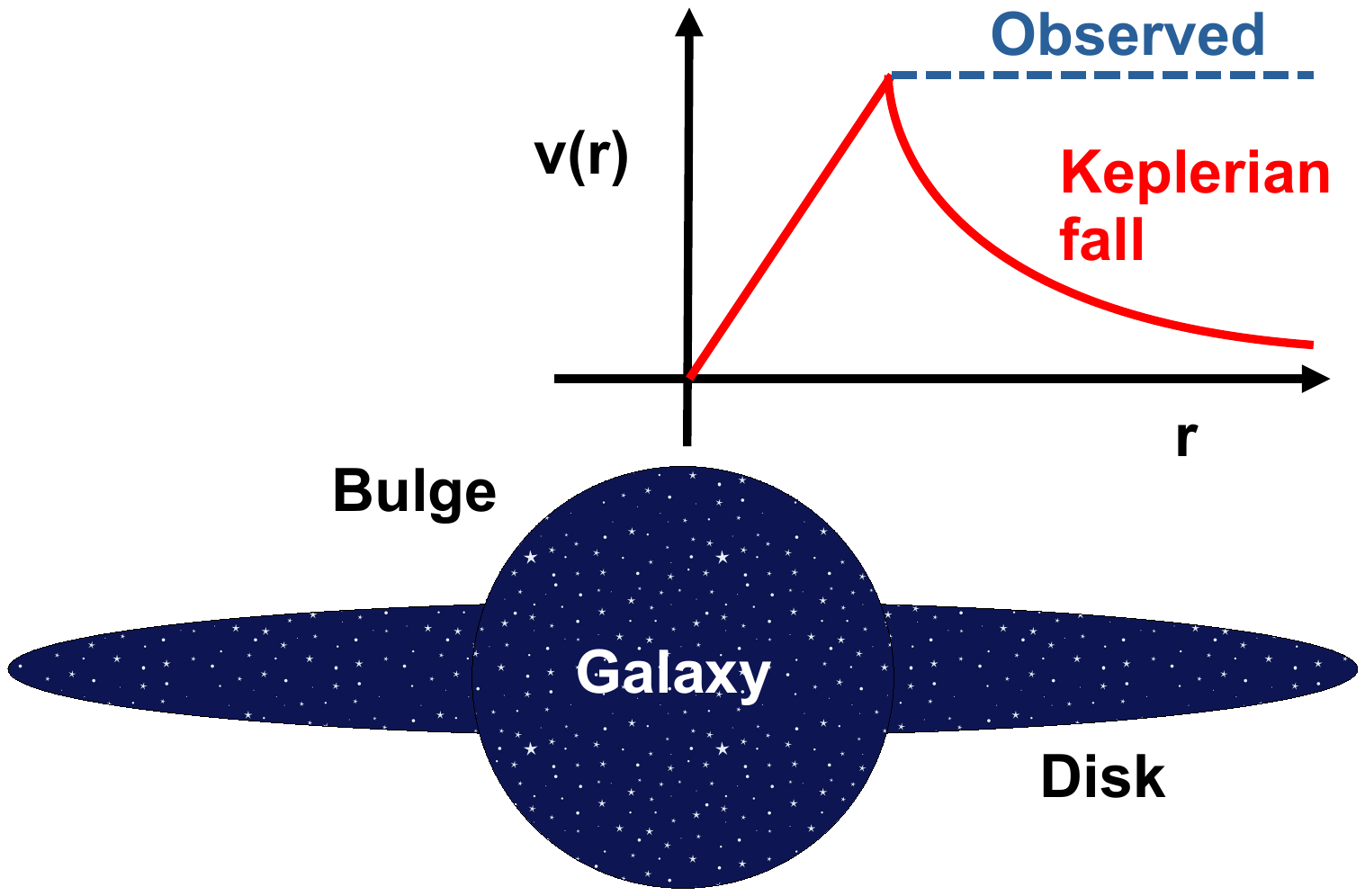}
    \caption{Schematic view of the expected Keplerian fall. The behavior of the rotation curve $v(r)$ (top) is shown both inside and outside the galaxy bulge, with its cross section displayed at the bottom.}
    \label{fig:gal}
\end{figure}

The discussion about the details and the status of the direct dark matter search will proceed as follows: In Sec.~\ref{sec:easy}, the reason, according to the authors, why the direct dark matter detection is so popular, and increasingly supported and financed is outlined; in Sec.~\ref{sec:det}, the basic ideas about the direct detection of dark matter are reviewed;
in  Sec.~\ref{sec:anal}, important and often undervalued aspects of the direct dark matter analysis are highlighted; in Secs.~\ref{sec:nai} and~\ref{sec:nob}, the case of both NaI-based and noble gas (xenon and argon) detectors are largely reviewed, respectively. Finally, in Sec.~\ref{sec:evol},  the evolution of dark matter search in the new millennium is critically reviewed and analyzed.

\section{A simple theory} \label{sec:easy}

The direct dark matter search has received an increasing consensus in the recent decades also thanks to its simplicity, and then to the (apparent) solidity of its foundations, invoking some principle invented in the context of theology as the \emph{Occam's razor.}
The possibility of such a detection is enclosed in  simple equations, understandable at the level of an average high school student. Three basic formulas, concerning the dark matter hypothesis and its possible detection, are hidden among a list of seemingly simple equations. Here is the list:
\begin{eqnarray}
\label{eq:1} & \nexists A : \aleph_0 < \#A < 2^{\aleph_0}  \\
\label{eq:2} & v = \sqrt{GM/r} \\
\label{eq:3} & \vec{\nabla} \cdot \vec{\mathcal{F}} = 0, \vec{\nabla} \times \vec{\mathcal{F}} = i \frac{\partial \vec{F}}{\partial t} \\
\label{eq:4} & \mathcal{H} u = -\frac{\partial v}{\partial t}, \mathcal{H} v = \frac{\partial u}{\partial t} \\
\label{eq:5} &E_A = \frac{4 M_A M_\chi}{(M_A + M_\chi)^2} E_\chi \\
\label{eq:6} &x^3 + x = 1 \\
\label{eq:7} & \mathcal{R} = n\sigma \Phi \\
\label{eq:8} & \int_\Omega \mathrm{d}\omega = \int_{\partial \Omega} \omega \\
\label{eq:9} & (i\hbar \gamma^\mu \partial_\mu  - m c) \psi = 0 \\
\label{eq:10} & e^{i\pi} + 1 = 0
\end{eqnarray}

An expert reader has immediately recognized that the relevant three equations are Eq.~(\ref{eq:2}), Eq.~(\ref{eq:5}) and Eq.~(\ref{eq:7}). In the following Subsections, an explanation for each of the three Equations is discussed\footnote{For the sake of completeness, the explanation why the rest of the listed equations look simple, but they are not, is hereby reported. The expression (\ref{eq:1}), both simple and deeply complicated, represents in symbols the Continuum Hypothesis, partially still the subject of discussion among modern mathematicians. The Eqs. in (\ref{eq:3}) represent the Maxwell equation in vacuum, just taking the electric and the magnetic field as the real and the imaginary part of $\vec{\mathcal{F}}$. Whereas, the Eq. (\ref{eq:4}) displays the Shr\"oedinger equation, but this time with two real functions $u$ and $v$! Equation (\ref{eq:6}) is really simple, but it has ``complex'' solutions, definitely not easy to be found if one ignores some algebraic technicalities. The Eq. (\ref{eq:8}), usually called the Stokes theorem within the $k$-form formalism, is probably the most beautiful theorem in advanced calculus, as it encloses the Fundamental Theorem of Calculus, the 3D Green and Stokes theorems, their 4D version in the Riemann space-time and so on. The Equation (\ref{eq:9}) is the famous Dirac equation, whose correct interpretation in Quantum Field Theory predicts the existence of antiparticles for Fermions. Finally, the Eq. (\ref{eq:10}), the Euler identity, often advertised by Richard Feynman for its beauty, is a complex number identity that is far more tricky than its apparent simplicity, as it involves the top five numbers in mathematics.}.

\subsection{Existence -- Eq.~(\ref{eq:2})}

A spiral galaxy contains about two thirds of its mass ($M$) in the galactic core (or bulge)~\cite{bib:sal1, bib:sal2}. Assuming for the latter a spherical shape with spherically symmetric density $\rho(r)$, the Gauss theorem predicts, outside the bulge, a gravitational pull $\propto M/r^2$, where $r$ is the distance from the center of the sphere and $M$ is the enclosed galactic mass (see Fig.~\ref{fig:gal}). More precisely, the profile of the visible matter depends on the galaxy type, but its specific realization does not invalidate this general argument.
This force provides a centripetal acceleration $\propto v^2/r$ for all objects gravitating around the galaxy center. Here one has implicitly assumed the Newtonian weak field approximation of General Relativity at the galactic scale and at the galactic speed (this point has been recently debated, see~\cite{bib:GRgaia} already anticipated above).  

The model described so far immediately implies that the velocity of stars as a function of $r$ should follow the so-called ``Keplerian fall" described by Eq.~(\ref{eq:2}). But this is not what is observed \cite{bib:rubin}: rotation curves always lie largely above the expected behavior coming from the independent quantification of the visible galactic mass from mass-to-luminosity ratio of stars and non-luminous (gaseous or solid) mass. Sometimes it is somewhat inaccurately said that observed rotation curves are \emph{flat}. Those curves are actually exceeding the Keplerian fall, but with a family of universal curves that depends on the visible size of the galaxy, upon which the ratio between visible and dark matter also depends~\cite{bib:sal1, bib:sal2}. The dark matter distributed in a spherically symmetric halo, present in the interstellar space inside each galaxy, fills this discrepancy between predictions and observations.

There is considerable debate about the profile of the radial dependence of the spherically symmetric matter distribution in the hypothetical dark matter halo, depending on whether it is inferred from numerical simulations of collision-less particle clustering or from observed rotation curve analysis.
The majority of the proposed profiles can be described by the generic formula:
\begin{equation} \label{eq:profs}
   \rho_s(r)=\frac{\rho_0}{\sum_{i=0}^3 a_i \left(\frac{r}{r_s}\right)^i},
%   \rho_s(r)=\frac{\rho_0}{ a_0  + a_1 \left(\frac{r}{r_s}\right)
% 	+ a_2 \left(\frac{r}{r_s}\right)^2  + a_3 \left(\frac{r}{r_s}\right)^3},
\end{equation}
where $\rho_0$ is a normalization constant, and $r_s$ is a characteristic size scale. According to the choice of the dimensionless coefficients $a_i$, the profile can be more ``cuspy'' (as in the Navarro-Frenk-White model~\cite{bib:nfw}) or more ``cored'' (as in the pseudo-isothermal or Burkert models~\cite{bib:burkert}); or something different, but anyway traceable by approximation to Eq.~(\ref{eq:profs}), as in the Einasto profile family~\cite{bib:einasto}.  

To conclude this Section, it is worth mentioning that, interestingly, the absence of the Keplerian fall has been recently questioned, at least from some interpretation of accurate observations of the Milky Way by GAIA~\cite{bib:gaia1, bib:gaia2}. The correctness and the implication of such results have yet to be validated and fully understood.

\subsection{Kinematics -- Eq.~(\ref{eq:5})}

A typical dark matter candidate $\chi$ of mass, say, $M_\chi=50$ GeV/c$^2$, hitting a target (visible) nucleus $A$ with mass of about $M_A=50$ GeV/c$^2$ in laboratories, has a typical velocity $v_\chi$ in the galaxy of the order of a few hundreds of km/s ($v_\chi \ll c$). As a consequence, the kinematics of the collision $\chi$--$A$ is not different from two  billiard balls with masses $M_\chi$ and $M_A$ hitting each other (in a non relativistic regime, actually $\beta =v_\chi/c \sim 10^{-3}$). The classical energy and momentum conservation leads to Eq.~(\ref{eq:5}), in case of linear collision, otherwise the decrease for the cosine of the scattering angle has to be included. The kinetic energy of the target recoil $E_A$ is a \emph{kinematic} fraction of the dark matter kinetic energy $E_\chi$. Based on the numbers provided above, calculations show that the expected recoil energy in a detector on Earth is of the order of $\lesssim 10$ keV, and so extremely difficult to detect, but not impossible with the present technologies.

\subsection{Rate -- Eq.~(\ref{eq:7})}

Lastly, Eq.~(\ref{eq:7}) represents the interaction rate $\mathcal{R}$ expected in a given detector on Earth, exposed to the dark matter wind caused by the Solar System's relative motion inside the Milky Way: the rate is given by the target density $n$ times the $\chi$--$A$ cross section per nucleon $\sigma$, typically assumed to be spin-independent (SI), and the dark matter flux $\Phi$. In detail, assuming a Maxwell distribution of the dark matter particle velocities inside the gravitational potential ``box'' of the Galaxy, and the experimental features of the detector, Eq.~(\ref{eq:7}) has to be integrated over the velocity distribution, taking into account the detector energy threshold. Then, all the data has to be combined with the experimental resolution, normalized to the energy quenching for nuclear recoils, and, finally, scaled according to the detector acceptance~\cite{bib:lewin}.
In the SI case, the full formula can be summarized as
\begin{equation} \label{eq:rate}
    \frac{d\mathcal{R}}{dE} = \frac{\rho_{\rm \chi} A^2 \sigma_{\rm}F^2(E)}{2M_{\chi} \mu^2} \int\limits_{v_{\rm min}}^{v_{\rm esc}} \frac{f(v, v_0)}{v} \mathrm{d}v \otimes  \mathcal{G}(E)
\end{equation}
Where $\sigma$ is the cross section per nucleon of the target $A$, $F(E)$ is the nuclear form factor, $\mu$ is the reduced mass between the target and the dark matter mass. Whereas, assuming the standard (galactic) halo model (SHM)~\cite{bib:halo1, bib:halo2},  $\rho_\chi=0.3$ GeV/cm$^3$ represents the local dark matter density in the Solar System, $v_{\rm min}$ the minimum delectable velocity (given the experimental energy threshold) and $v_0=220$ km/s the circular rotation velocity and $v_{\rm esc}=544$ km/s is the Milky Way escape velocity, taken as the cut-off for the Maxwellian velocity distribution. The Sun velocity is $v_\odot=232$ km/s. Those parameters have been slightly updated recently, but those small variations go beyond the basic purpose of this article.
It is worth noticing that the notion of SHM could be criticized as non-realistic, and one can imagine a vast zoology of monster halos with local density anisotropies and streams: this can change a little the result interpretation (as the rate normalization can change), but cannot create specific detection anomalies out of nothing.
Finally, the convolution, $\otimes$, with the energy dependent function $\mathcal{G}(E)$ symbolically includes the experimental features, such as resolution, nuclear quenching and acceptance in the region of interest.
\begin{figure}
    \centering
    \includegraphics[width=\columnwidth]{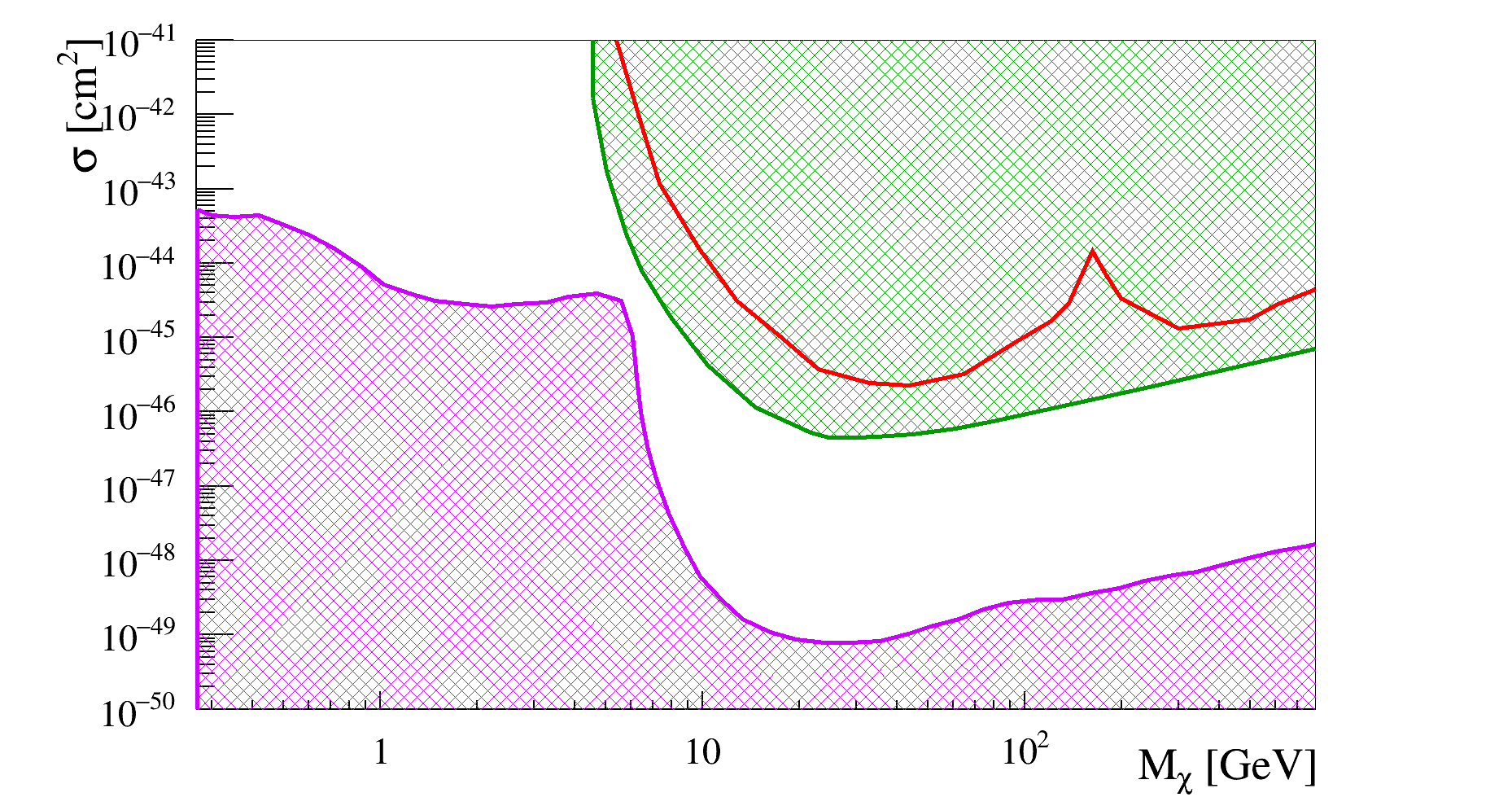}
    \caption{The greenish region represents the typical excluded region of dark matter experiment with null result in the SI $\sigma$--$M_\chi$ plot. The violet region represents the so-called neutrino floor. Finally, the red curve, with a cusp, represents a limit in case of annual modulation analysis, see Sec.~\ref{sec:anal}.}
    \label{fig:floor}
\end{figure}

The Equation~(\ref{eq:rate}) can be integrated over the experimental energy window and made explicit as $\sigma_{\rm SI}(M_\chi)$. An experimental limit, \emph{e.g.,} the absence of events at 90\%CL,  typically looks like an asymmetric hyperbole branch, as depicted in Fig.~\ref{fig:floor} (greenish region): the branch on the left represents the experimental threshold wall, instead the branch on the right corresponds to the loss of sensitivity due to the reduced density of targets for heavier dark matter masses. An experiment has typically the maximal sensitivity for $M_\chi \simeq M_A$, corresponding to the minimum of the green curve. The violet region in the Figure corresponds to the so-called \emph{neutrino floor}, \emph{i.e.} the region in which neutrinos coming from Sun, atmosphere and diffused supernova background gives a nuclear recoil from coherent scattering, similar to the expected equivalent dark matter interaction with a given mass. This actual experimental limitation can be overtaken only by future experiments exploiting the ``directionality'' of the dark matter wind, \emph{i.e.} the preferred direction along the galactic plane, due to the relative motion of the Sun with respect to the galactic center, against the neutrino background, assumed as uniform and isotropic. This particular aspect will be further discussed in Sec.~\ref{sec:evol}.

\begin{figure}[th!]
    \centering
    \includegraphics[width=\columnwidth]{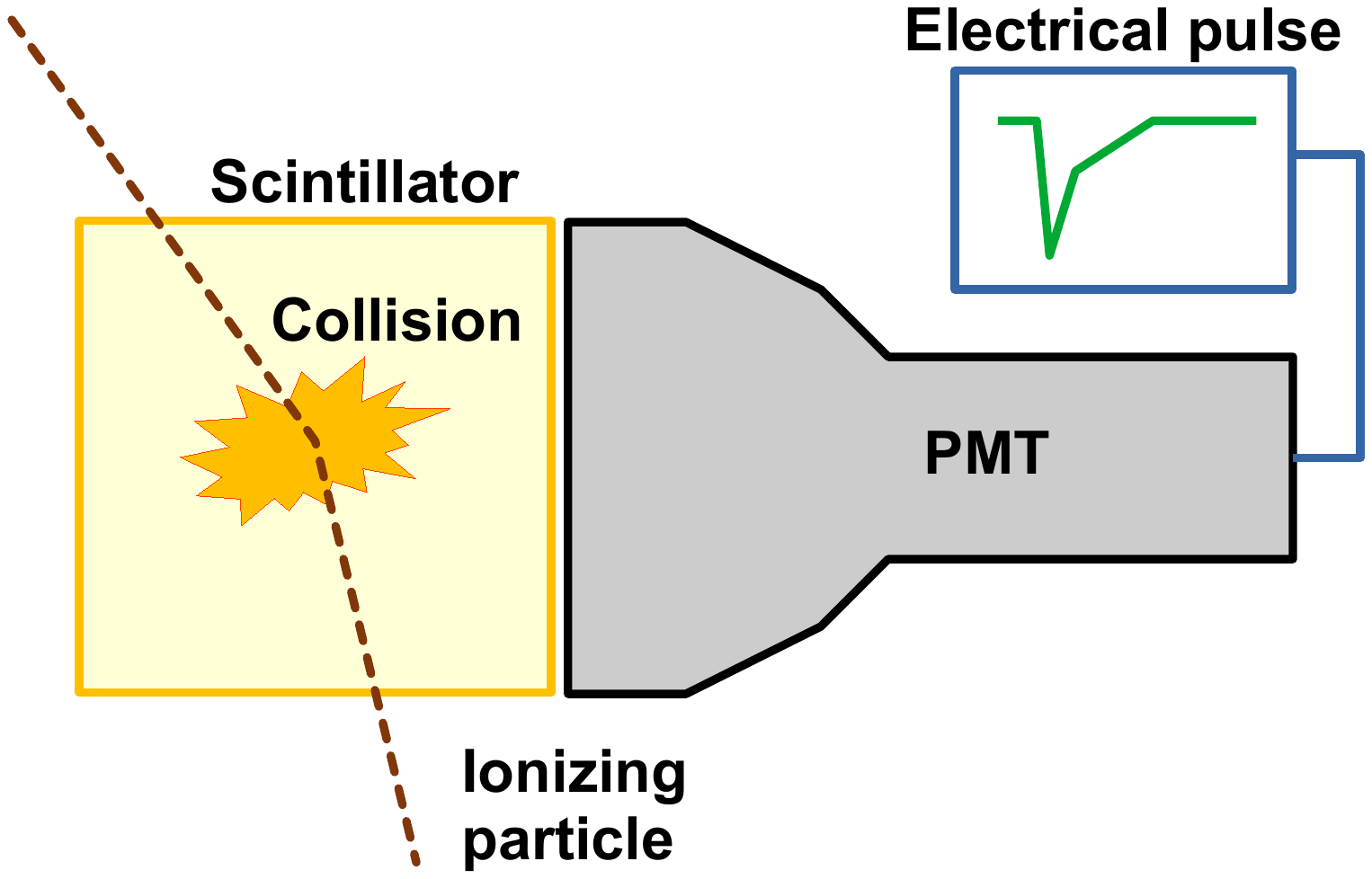}
    \caption{Example of a simple detector made of a scintillating crystals optically coupled to a PMT. The scintillation signal is converted into a series of photo-electrons producing an electrical pulse related to the time distribution of the scintillation light emission.}
    \label{fig:pmt}
\end{figure}

\section{Detectors} \label{sec:det}

A grand piano is constructed from durable and raw materials. Indeed, it is basically made of hard substances, such as wood, cast iron, steel and felt. A single string holds a tension of about 80 kg and many of its 88 keys own 2 or 3 strings, for a total tension of 15000 kilogram-force. However, only after a wise assembly and a proper (fine) tuning, this terrific music instrument can produce a very soft and profound sound, which one can appreciate for example in the famous Chopin's Nocturnes. Similarly, a particle detector is made of hard and raw stuff as well: from metals, crystals or liquids and electronics, one can get an answer about fundamental question in particle physics and cosmology, but this process is not that straightforward, and a poor knowledge of its functioning can deceive the smart experimentalist, and the brilliant theoretician who wants to jump too quickly to the conclusion.

Consider a simple example. A typical particle detector is made of a scintillating target (solid or liquid) coupled with a light sensor, for example a photo-multiplier tube (PMT)~\cite{bib:knoll}. When an ionizing particle hits the target, a given number of electromagnetic field quanta are excited, and part of them (depending on detection efficiency) collapse to form what is known as a ``photo-electron'' on the light detector. The corresponding electric pulse output is then shaped and amplified and eventually converted into a binary number, and finally stored on a computer hard drive. The corresponding data are retrieved and analyzed by numerical algorithms or passed to some artificial intelligence black box, see Fig.~\ref{fig:pmt}.

A real detector is, of course, sensitive to internal and external radioactive backgrounds, as well as hypothetical dark matter particles, whose foreseen amount plays a crucial role in the goodness of the proposed experimental setup. Furthermore, the electronic (non physical) noise, intrinsic and/or picked up from the environment, can mimic, to some extent, the signal produced by particle interactions. As a matter of fact, only a very deep knowledge of all those effects described above, can enable a good investigation. And sometimes, as it happens in most of the experiments, part of those effects are not known \emph{a priori} and can be addressed properly only after a lot of years of calibrations, analysis and hardware improvements. Even a seemingly simple device like a PMT can produce a wide variety of noise pulse signals that might not be distinguishable by a basic classifier, and requires a deep characterization, sometimes using novel techniques based on multidimensional mathematical algorithms such as multivariate likelihood ratios or support vector machines, or even nonlinear methods based on boosted decision trees or multi-layer perceptrons, and machine learning in general.

\begin{figure}
    \centering
    \includegraphics[width=\columnwidth]{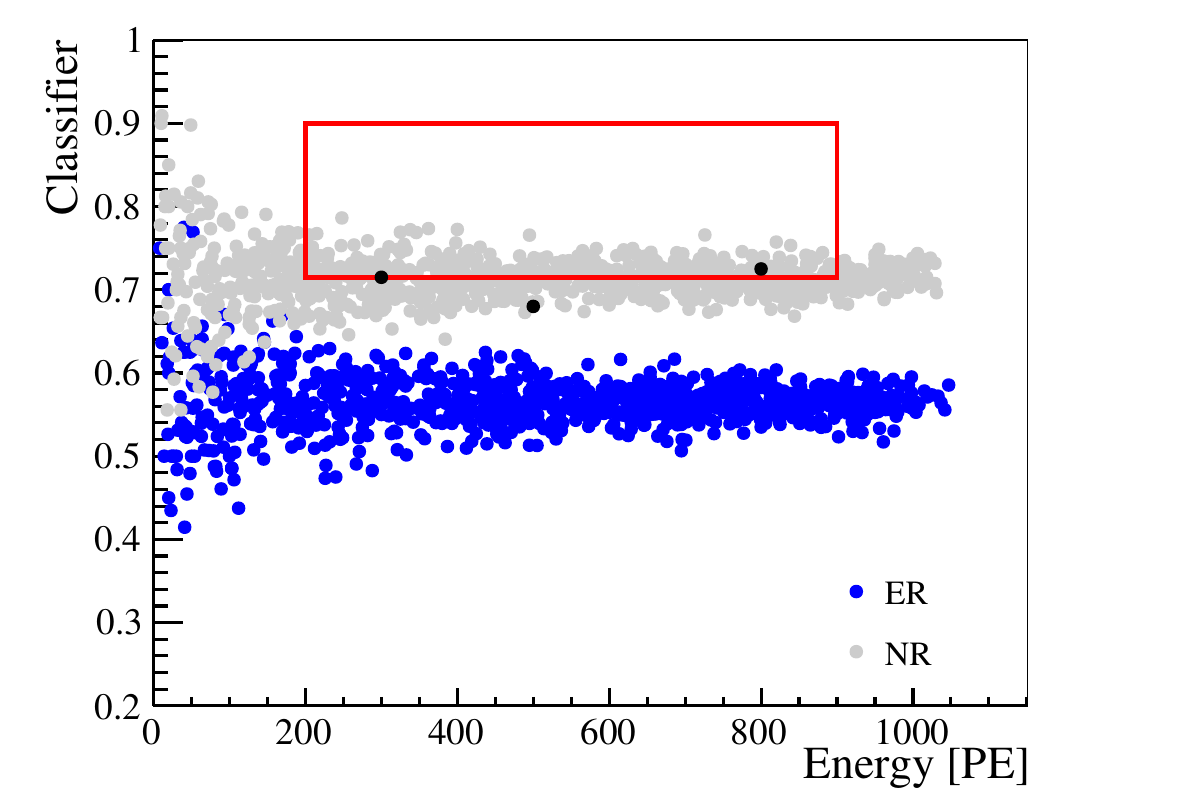}
    \caption{Classifier distribution for NR (gray) from calibration and ER (blue) from a physics data collection. The acceptance region (red box) for NR candidates can be defined e.g. as the median of the NR distribution. The observed three events in the NR band are to be interpreted according to event/noise understanding and expected physical background.}
    \label{fig:class}
\end{figure}

Finally, the optimal detector has some properties capable of discriminating, not only between signal and noise, but also between the physical characteristics of the primary interaction, as between electron and nuclear recoils (ER--NR, hereafter), especially useful for addressing the nature of a possible dark matter candidate. In the example mentioned above, the discrimination could be enabled by the time distribution of the scintillation light. Typically,  a classifier, $\emph{i.e.}$ a parameter defined through  $\emph{e.g.}$ a likelihood ratio or an artificial neural network, shows a characteristic distribution depending in general on the particle energy and exhibiting a partial overlap (inefficiency) in the region of interest. Usually, after a deep training with known sources, an acceptance region, in which the dark matter candidate is expected, is defined (See Fig.~\ref{fig:class} as an example). If a statistically significant group of events emerges over the expected background, one can reasonably claim evidence of new physics, which should later be confirmed by other equally sensitive experiments or complementary techniques.

Dark matter detectors typically exploit various phenomena to develop classifiers. An ionizing particle hitting a material can, in general, produce scintillation, ionization or phonon excitation. Many detectors, depending on state and on the temperature, can utilize one or more signals, offering broader capabilities in particle discrimination~\cite{bib:schumann}.

The high level standard of low background detection requires the choice of highly radio-pure target materials to be operated in shielded and, possibly, actively vetoed detectors, located in underground laboratories, far enough from the atmospheric muons and cosmogenic-induced radiations. These materials are usually not available on the market and require a long and accurate R\&D program, sometimes with no predictable outcome.  

Lastly, an important experimental aspect to consider is nuclear quenching. A NR indeed releases only a portion of its actual kinetic energy due to non-radiative excitation. For that reason, the observable energy is less than the one released by an equivalent ER. This relative ratio is important for the final interpretation of the experimental result, for this reason results are usually presented in electron-equivalent energy (\emph{e.g.} keVee) and the quenching factor is given independently by dedicated calibrations with neutrons.

Besides the counting of outliers in the expected acceptance region, direct dark matter detection in underground laboratories can also be pursued through the detection of annual modulation. This signal arises from the relative motion of the Earth around the Sun with respect to the center of the Galaxy. In addition, tracking detectors sensitive to the directionality of the dark matter can be employed.

With these basic concepts in mind, one can now move to the present experimental situation.

\section{Analysis} \label{sec:anal}

Imagine that one gets a few (unsuspected) points in the square box of Fig.~\ref{fig:class}. Can one claim for the dark matter discovery based on that? A wise response might be that the number and the distribution (\emph{e.g.} in energy and space) of the expected background has to be declared in advance. For example, if an experiment observes 4 events out of 2 (expected), one may argue that the Poissonian fluctuation of 2 can likely give 4 in a reasonable number of cases. But if one gets 8 out of 2, the story becomes more intriguing, as the fluctuation of 2 can hardly return 8.

\subsection{Which statistics?}

The way in which these naive words ``likely" or ``hardly" are converted into some quantitative parameters is not universally accepted by a common procedure, and there are multiple approaches based on different statistical interpretations, often with heated debates, like the ones between  Frequentists (\emph{e.g.} Feldman-Cousin~\cite{bib:FC}) or Bayesians~\cite{bib:bayes} statisticians.

Of course this is a fundamental and longstanding controversial debate that cannot be solved for sure here and sometimes the solution is not unique, but depends on the specific situation.  Therefore, for an experimentalist very often it is more convenient to quote results (whether it is a measurement or a limit) in more than one approach, just to delegate some possible controversial matters to others.

The real challenge might be how to properly combine results coming from different approaches. However, it is also true that as long as results are just limits the process itself is not really harmful, and the nuances in results stemming from different philosophies are basically hidden by the line width in the $\sigma$--$M_\chi$ exclusion plot. For real positive results, the problem could be more delicate.

\subsection{How many \texorpdfstring{$\upsigma$}{sigma}'s?}

If some statistical procedure is assumed, it becomes important to show how many sigmas (actually the $p$-value over the background fluctuation) are necessary to claim for a discovery. Among physicists, there is a common (questionable) practice to associate a naive meaning to a certain number of sigmas, such as:
\emph{mild indication} ($1\sigma$), \emph{indication} ($2\sigma$),
\emph{evidence} ($3\sigma$), \emph{never quote that, it's bad luck!} ($4\sigma$) and \emph{discovery} ($5\sigma$). The real problem here is that all of them are only indicators: forgotten systematic uncertainties and mistakes are always possible and may unexpectedly arise, and unfortunately there is plenty of such examples in literature.

Besides these folkloristic topics,  to claim a scientific discovery, the way is much harder, as it is explained in the following Subsection.

\subsection{No background}

Given that background fluctuations can deceive experimentalists, it would ideally be preferable not to have any background at all. In other words, when planning a detector, it is ideal for the expected background events to be significantly fewer than the expected signal to avoid being misled by fluctuations. This has created the fashion of having high-sounding names as \emph{background-free} experiments, or even \emph{zero-background} experiments.

What is the difference between ``Cola Light" and ``Cola Zero"? In reality, there is no difference: both contain chemical sweeteners instead of sugar, but in marketing names really matter. Some people might avoid a drink labeled as ``Light", but would choose one labeled ``Zero", as if it is not for everyone facing the zero with no fear. The same concept applies to the background in dark matter experiments: it is primarily marketing. The honest name should be ``low background", where the adjective ``low"  should go with some objective quantifier.

In a popular, and absolutely interesting, review on neutrinoless double beta decay, a footnote reads~\cite{bib:viss}:
\begin{quote}
 ``For a truly zero background experiment, one event is enough to
claim a discovery. In a similar fashion, encountering a unicorn is enough to claim its existence, provided that we have a template of a unicorn to which to compare the observed candidate.''   
\end{quote}
When the authors asked \textsc{GPT-4}\cite{bib:gpt4} to comment on this statement, its response was notably insightful. Here is its reply:
\begin{quote}
``A single observation of an event can suggest that something exists, but in a scientific context, such conclusions are usually considered weak evidence. Scientists value repeatability and reliability of data, meaning that multiple, consistent, and independent observations are necessary to establish existence.

In the case of a unicorn sighting, a single observation would likely not be considered sufficient proof of their existence. Scientists would need to examine the evidence, conduct additional observations, and demonstrate that these observations are indeed consistent and repeatable. For rare or extraordinary phenomena like unicorns, scientists may be more cautious and require even more evidence to accept their existence.

In summary, a single observation can lead to the belief that something exists, but scientists typically rely on repeatable, reliable, and independent data. A lone unicorn sighting would likely be insufficient, and further research and observations would be needed to confirm their existence.''    
\end{quote}

If one ignores this crispy reasoning, one can be easily fooled and forget the basic principles of good science. Indeed, continuing with the metaphor, the unicorn is quite an unfortunate example, not only for our prior about its existence, but also because there is plenty of alicorns (unicorn's horns) in museums, coming from the Middle age~\cite{bib:narval}, sometimes made actually of bone. Yes, but a lot of people, even today as in the Middle Age, ignore the existence of narwhals.

To conclude, what one sees in a detector is not a unicorn anyway, if for unicorn one means something completely weird and different from ordinary stuff. In a real particle detector, what one records is nothing else than a pure electronic signal, digitized and analyzed as a discrete waveform, that resembles either physical events and/or noise artifacts more often than expected.

\subsection{Annual modulation}

\begin{figure}
    \centering
    \includegraphics[width=\columnwidth]{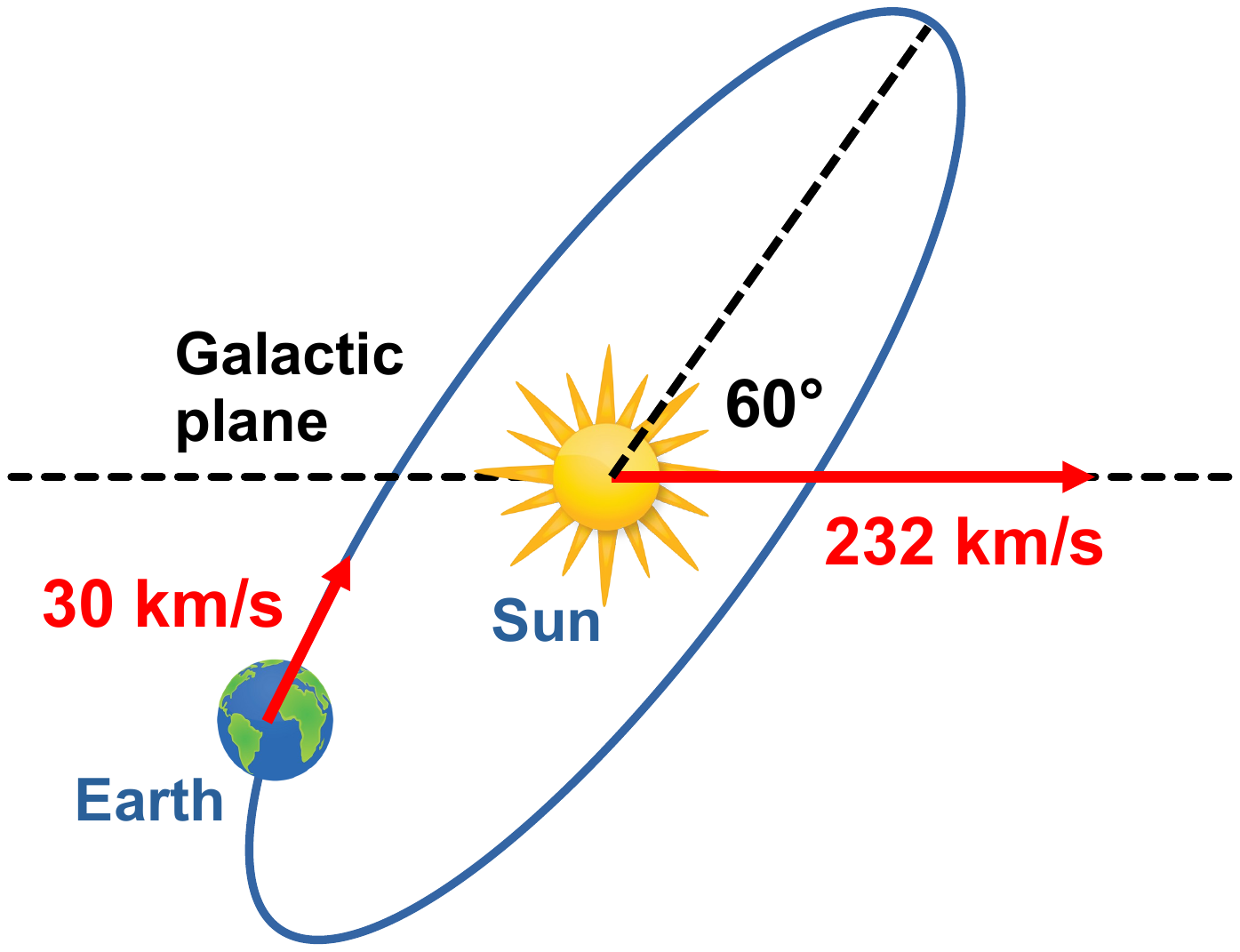}
    \caption{Diagram of the Earth and Sun velocity vectors with respect to the galactic plane.}
    \label{fig:ssys}
\end{figure}

A ``model-independent" approach, which in principle can ignore the nature of the $A$--$\chi$ interaction, involves detecting an annual modulation signal from a shielded detector over a prolonged (multi-annual) exposure period. In this case, the ER--NR discrimination  is not necessary, as one is interested only in the typical signature of such a signal.
The Earth, indeed, revolves around the Sun at 30 km/s while the Solar System, tilted of 60° with respect to the galactic plane, moves altogether around the center of the Galaxy at about 232 km/s, as depicted in Fig.~\ref{fig:ssys}. As a result of this simple geometry, the expected signal is
\begin{equation} \label{eq:mod}
   \mathcal{R}(t) = \mathcal{R}_0(t) + \mathcal{S}_m \cos(\omega(t-t_0)),
\end{equation}
\begin{figure}[ht!]
	\centering
	\includegraphics[width=\columnwidth]{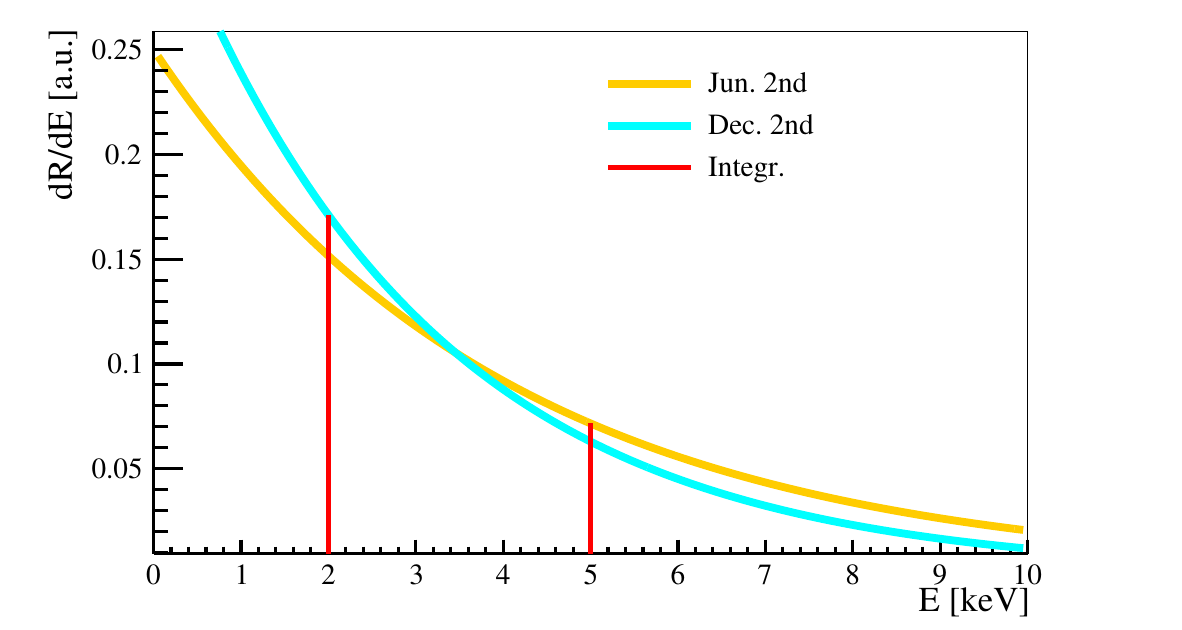}
	\caption{Two limit cases of distortion: June 2nd (orange), with maximum relative velocity of the Earth with respect to the dark matter halo, and the opposite case on December 2nd (cyan). Finally, the red lines mark a possible experimental integration interval.}
	\label{fig:inversion}
\end{figure}
\begin{figure*}[ht!]
    \centering
    \includegraphics[width=0.9\textwidth]{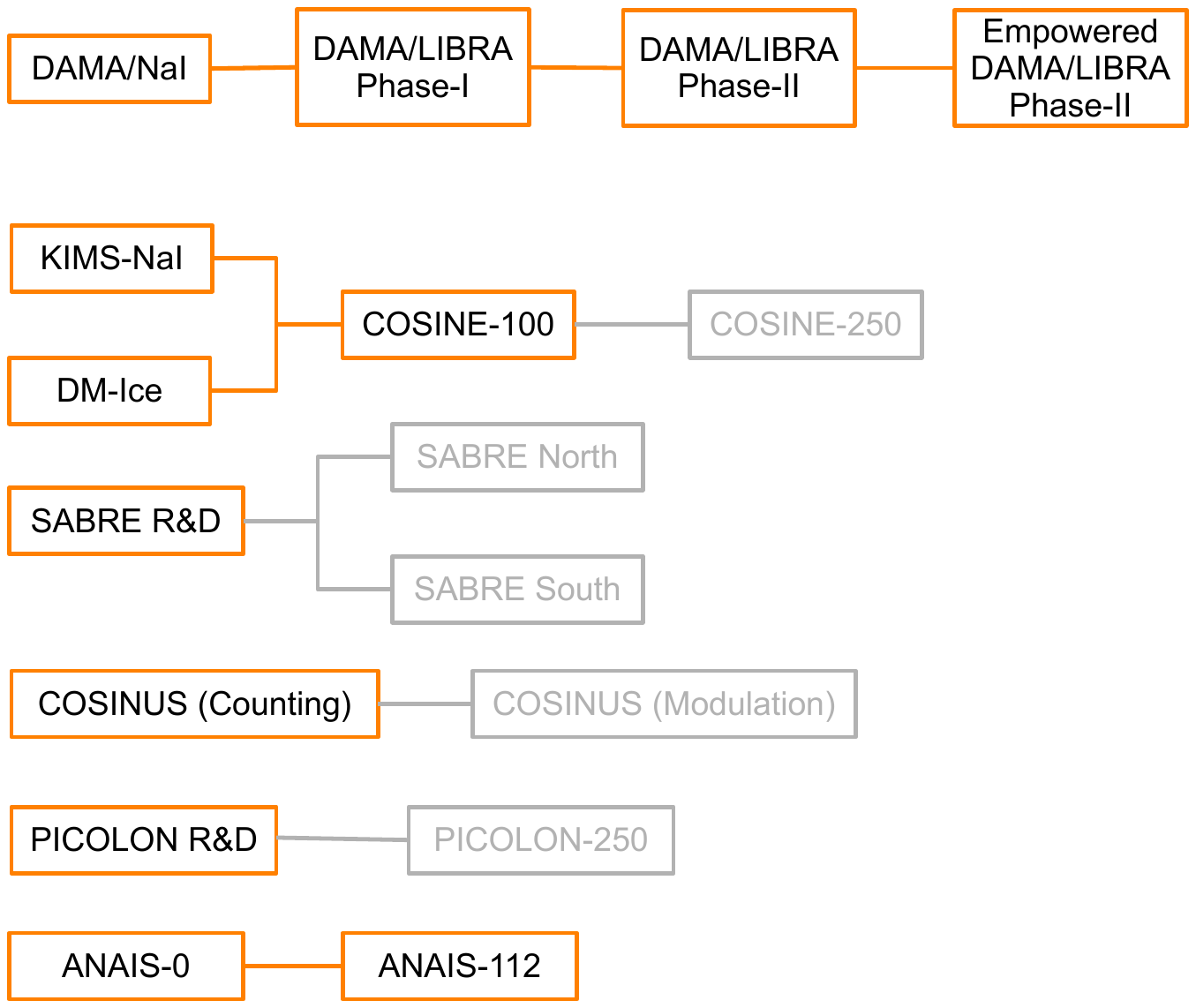}
    \caption{Genealogy of the NaI-based detector. Besides COSINUS, which is planning to deploy cryogenic crystal with ER--NR discrimination, all the others are using the model-independent approach exploiting the annual modulation expected signal.}
    \label{fig:gen}
\end{figure*}

Where $\mathcal{R}_0(t) = \mathcal{S}_0 + \mathcal{B}(t)$ is the total trend given by the sum of a constant part $S_0$, due to the \emph{unmodulated} dark matter component, and the detector background $\mathcal{B}(t)$ that in principle can depend upon time, as \emph{e.g.} in presence of radioactive decaying contamination, or time varying cosmogenic background; $\mathcal{S}_m$ is the modulation amplitude of the (expected) dark matter signal, basically given by the relative speed of the Earth with respect to the local dark matter velocity distribution in the Galaxy, and therefore of the order $30/232*\cos( 60°) \simeq 6.5\%$
with respect to $\mathcal{S}_0$; $\omega$ is the annual angular velocity corresponding to $2\pi/(365\, {\rm d})\simeq 0.0172$ rad$^{-1}$;  finally, $t_0$ is the time (phase) corresponding to the maximum rate, \emph{i.e.} to the date in which Earth and Sun go towards the same direction (on June 2nd). If one builds up a radio-pure and shielded setup, acquires data for some years and performs a temporal regression analysis in which $S_m$, $\omega$ and $\phi$ are free parameters returned simultaneously in the expected ranges, one can in principle claim that such a signal is compatible with the presence of a diffused dark matter gas in the interstellar space at the Earth distance from the Galaxy center. Indeed, Freese et al.~\cite{bib:freese} clearly states:
\begin{quote}
    ``We argue that a modulation can itself be used as primary means of detecting WIMPs [more in general dark matter (A.N.)] (rather than as a confirmation) even in presence of a large background.''
\end{quote}
This sentence is \emph{for common sense} correct but actually \emph{theoretical} and \emph{practically} wrong. First of all, it corresponds to the material implication ``{\bf IF} the dark matter exists {\bf THEN} a modulation is visible", \emph{i.e.}
\begin{equation}
\textsc{dark matter} \rightarrow \textsc{modulation},    
\end{equation}
and so, looking at its truth table, the fact that one can see modulation even if dark matter does not exist is still a valid possibility: one can argue for example, that a time-varying background related to some possible seasonal (o seasonal-induced) signal is still possible~\cite{bib:mu1, bib:mu2};
moreover, a time-varying $\mathcal{B}(t)$ background, if not properly accounted for, can bias the final results~\cite{bib:rossi}, as it is discussed later. Then a robust modulation analysis has to show the consistency of all terms of Eq.~\ref{eq:mod}, no one excluded, at the very least. As a further example, it is also worth mentioning that the literature is full of apparent violations of the exponential law with annual modulation components in radioactive decays,
opportunely criticized, see~\cite{bib:pelczar} and Refs. therein.

Finally, in case of annual modulation the limit in Fig.~\ref{fig:floor} has a cusp around the minimum (red curve). This is an artifact of integrating each time bin over a limited energy window. It is possible, indeed, that the distribution of the target recoil spectrum is distorted between June 2nd ad December 2nd in such a way that the shape changes, but the total area not: in this case there is a specific dark matter mass for which the sensitivity is lost, see Fig.~\ref{fig:inversion}. The same accurate analysis also leads to the so-called \emph{phase inversion} phenomenon, happening in the very low energy region, that should be present in the experimental data, but sometimes not visible because of a too high experimental threshold with respect to the inversion point that depends also on the specific dark matter mass.

\subsection{Blind analysis}

The blind analysis, consisting for example in closing, completely or in part, the red box of Fig.~\ref{fig:class}, is a good practice, adopted by other disciplines such as medicine, to avoid biases in the analysis, especially in low-rate critical conditions. A collaboration that decides to apply this practice, usually closes the box for physical data taking~\cite{bib:xent, bib:ds50last}, training the event reconstruction and the selection criteria only on a subset of data, possibly not used in the final analysis. The collaboration decides to freeze the data-set, and eventually to open the box. The scenario is significantly different when the ``un-blinding" is done in public with journalists.
With a few exceptions~\cite{bib:times}, this is usually done
behind closed doors instead, and nobody knows what happens inside. In this case one can only rely on the honesty and professionalism of colleagues.

\subsection{Data sharing}

It is worth citing what Wikipedia~\cite{bib:wiki} says about  the important item of the ``data sharing'':
\begin{quote}
``When additional information is needed before a study can be reproduced, the author of the study might be asked to provide it. They might provide it, or if the author refuses to share data, appeals can be made to the journal editors who published the study or to the institution which funded the research."
\end{quote}

Raw data are usually not understandable for non-experts. However, all physically reconstructed events and procedures must be available. The dark matter community should, sooner or later, converge towards an open data policy, especially if large amounts of funding are going to be spent on reproducing some experiments.

\section{The NaI case} \label{sec:nai}

Sodium iodide crystals doped with thallium NaI(Tl) are largely used as particle detectors exploiting their scintillation properties. The light produced when an ionizing particle hits the crystal is usually detected by PMTs, based on a very well known and consolidated technology, which nowadays can be easily manufactured with highly radio-pure materials~\cite{bib:xent}.

\begin{figure*}[ht!]
    \centering
    \includegraphics[width=\textwidth]{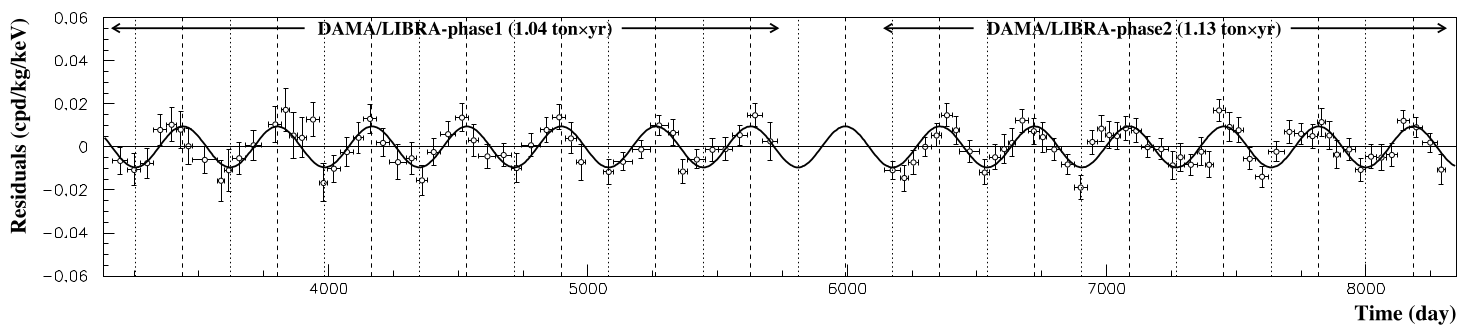}
    \caption{Experimental residual rate of the single-hit scintillation events measured by DAMA/LIBRA Phase-I and DAMA/LIBRA Phase-II in the (2–6) keV
energy intervals as a function of the time. The superimposed curve in Eq.~\ref{eq:mod},
with  period fixed to one year, phase fixed to 152 days (June 2nd) and amplitude
equal to the central value obtained by best fit on the data~\cite{bib:dama}.}
\label{fig:dama}
\end{figure*}

The possibility of using these crystals for dark matter direct detection was first explored by the DAMA collaboration (see~\cite{bib:dama} and  Refs. therein for a detailed review). DAMA, since its smaller version, has been detecting a modulation signal compatible with period and phase with the one expected by the presence of the dark matter halo in the Milky Way, rejecting the no-oscillation hypothesis at $12.9\sigma$ (20 annual cycles) in the 2--6 keVee energy interval. These interesting results, already released in its first and less massive version
at the turn of the new millennium, has become a media event, and has pushed, to some extent, the direct dark matter search in the first decade of 2000s, as it will be detailed in Sec.~\ref{sec:evol}. The signal detected by DAMA is significantly incompatible, in the framework of the SI interaction with SHM, with the absence of a corresponding signature in detectors with higher sensitivity, as the ones based on noble gases that will be discussed in Sec.~\ref{sec:nob}, and with other detectors exploiting a big variety of different techniques, briefly reviewed in Sec.~\ref{sec:other}. In other words and in practical terms, the DAMA signal is so intense that the same WIMP-like interactions should be visible in a cup of liquid xenon. Instead, xenon-based detectors,  as discussed later, are currently operating multi-ton targets with no results.
The remote explanation that the sodium and iodine nuclei have some ``special feature'', not met by other target nuclei, has anyway strongly motivated the scientific community to reproduce the DAMA experiment using similar NaI crystals independently, as it will be discussed later.

Figure~\ref{fig:gen} shows approximately the genealogy of experiments born after DAMA to accomplish this goal. At the moment, COSINE-100~\cite{bib:cosine} (originated by the merging of KIMS-NaI~\cite{bib:kims} and DM-Ice~\cite{bib:dm-ice}, and ANAIS-112~\cite{bib:anais} are the only two experiments already taking data for dark matter search. SABRE~\cite{bib:sabre} and PICOLON~\cite{bib:pico} are taking data in a R\&D stage, to basically prove the detection principle and quantify the intrinsic contamination. COSINUS~\cite{bib:cosinus} is the only detector using NaI crystals with a bolometric technique, and thus exploiting both scintillation and temperature signals for ER--NR discrimination. After some preliminary results on small crystal samples~\cite{bib:cos2}, the COSINUS collaboration is building a bigger and complete setup. In the following, the major experiments in this genealogy are discussed.

Finally, the interpretation of the NaI iodide results in terms of SI solution on the $\sigma$--$M_\chi$  parameter space could depend on the quenching factor for both Na and I nuclei separately, that could in principle depend on the specific crystal. There is a lot of debate on this issue and a lot of controversial measurement of those important parameters~\cite{bib:quench}.

\subsection{The DAMA experiment}

The impressive radiopurity of the DAMA crystals, as low as one count per day per kg per keV (cpd/kg/keV), with a threshold of $1\div 2$ keVee, made it possible to reach a sensitivity on the $A$--$\chi$ cross section better than $10^{-42}$
cm$^2$ for the model independent annual modulation in the energy region $\lesssim 10$ keV. However, the typical NaI(Tl) light yield of the order of $\sim 10$ PE/keV does not allow one for statistically sensible ER--NR discrimination. The DAMA collaboration had been operating
a 100 kg detector in the early phase called DAMA/NaI. The target was replaced with 250 kg of high purity NaI crystals in the subsequent Phase-I and Phase-II. In the latter, the energy threshold has been lowered from 2 keV to 1 keV. The DAMA collaboration is currently operating an empowered Phase-II with threshold as low as 0.5 keV with the aim of adding an further extreme low energy point in the analysis. This region is extremely important as it could show the phase inversion described above, not yet present in the recoil spectrum published so far by the DAMA collaboration.

The modulation measured by DAMA is $\mathcal{S}_m \simeq 0.01$ cpd/kg/keVee in the 2--6 keVee energy interval, extracted through a time fit of the residual single-hit (non coincident) rate as reported in Fig.~\ref{fig:dama} of Ref.~\cite{bib:dama}.

If one considers that this quantity represents only 6.5\% of the total rate detected in the crystals, one can naively assume, by scaling, that in the same energy window the unmodulated component of the total rate is about $\mathcal{S}_0\simeq 0.15$ cpd/kg/keVee, and then smaller than the total rate, \emph{i.e.} $\lesssim 1$ cpd/kg/keV.

As from Fig.~\ref{fig:dama}, the modulation reported by DAMA is extracted on the experimental residual rate of the single scintillation hits, \emph{i.e.} $\mathcal{S}(t) - \hat{\mathcal{S}}$(t) where $\hat{\mathcal{S}}(t)$ is the \emph{detrend} function. The detrend function used by DAMA is a piecewise function made with the average annual cycles of the total rate for each crystal. It is worth noticing that this method applies without bias, if and only if, the total rate of single-hit is constant; in all other cases, in which there is an explicit dependence upon time of the total rate, amplitude and phase are biased as described in~\cite{bib:rossi}. Basic signal processing theory, indeed, warns about the fact that injecting a periodical manipulation in time series will make the injected frequency itself appear in the final periodogram.
The total and explicit rate $\mathcal{S}(t)$ is never reported by the collaborations (see always~\cite{bib:dama} and Refs. therein), even if, from the same publications, it is evident that this rate is different in the two experimental phases, is time dependent because of the presence of decaying contaminants, and is presumably affected by many discontinuities because of hardware operations.

One is not saying that the DAMA signal is completely artifact from an incorrect analysis, but one is only suggesting that the result in phase and amplitude could be biased, and consequently the interpretation in terms of dark matter not exactly correct.

\subsection{Reproducing DAMA}

At present, SABRE is the only R\&D project that was able to manufacture high-purity crystals with a counting rate close to 1 keV/kg/keV comparable with DAMA~\cite{bib:frank}, and plans to deploy two different detectors in both Earth's hemispheres, to factor out all possible systematics due to unaccounted seasonal effects. PICOLON R\&D also plans to deploy a target of 250 kg of NaI crystals in forthcoming years. Finally, the COSINUS collaboration has proved that the combined scintillation and temperature signals in bolometric usage of the NaI can feature the ER--NR separation, and it is building the first real experiment for a counting experiment with a relatively small crystal. Furthermore, it plans, after the first data-taking, to increase the target and exploit the annual modulation analysis as well.   

In summary, the NaI case is not solved yet. The experiments trying to reproduce DAMA have not yet definitively clarified this debated result, therefore the funding agencies should strongly support all the attempts to independently clarify the question as soon as possible, with substantial resources and manpower. If there is a physical explanation for the DAMA signal, this explanation would be anyway highly interesting.  

\section{Noble gases} \label{sec:nob}

At present, noble gas based experimental setups are the most promising dark matter detectors in terms of sensitivity in a wide dark matter mass range, from 1 GeV/c$^2$ to 1 TeV/c$^2$~\cite{bib:xent, bib:ds50last, bib:deap, bib:panda4, bib:lz}, and recently even for masses lower than 1 GeV/c$^2$~\cite{bib:dslow, bib:xent}. Liquid targets made of these special elements, can reach high level of radiopurity and, thanks to their scalability, in terms of target mass, represent a realistic technology and a candidate for the ultimate experiment in dark matter search, capable of reaching the optimal sensitivity as will be discussed in Sec.~\ref{sec:evol}.

Among the known noble elements, argon and xenon are the sole gases permitting a feasible realization in terms of reliable technology and sustainable costs. Both gases can be exploited in single phase (liquid) detectors or double phase (liquid and gas) time projection chambers (TPCs).

Table~\ref{tab:arxe} summarizes the main properties of the two noble elements, making the two technologies complementary in terms of \emph{pros} and \emph{cons}. All details will be discussed in the two following subsections dedicated to xenon and argon, respectively.

Figure~\ref{fig:gen} sketches the genealogy of both technologies, showing weak (dashed) and strong (solid) relationships in terms of collaborators and/or merging of the corresponding experimental groups, while the gray block represents future projects.
Basically the two technologies have a common origin in large noble gas neutrino detectors,  as the ICARUS project~\cite{bib:icarus, bib:pvt}. Two distinct branches originated from some preliminary R\&D projects: the argon and xenon communities, even if this nomenclature is not formally shared among all collaborators. The genealogy includes neither the DAMA/Xe project~\cite{bib:damaxe} , a single-phase small scintillation not upgraded by the collaboration, nor XMASS~\cite{bib:xmass}, another single-phase xenon-based detector that has anyway set a fair limit on WIMP-like dark matter.  
Some of the experimental aspects of the two approaches will be discussed below.

\begin{figure*}[ht!]
    \centering
    \includegraphics[width=0.9\textwidth]{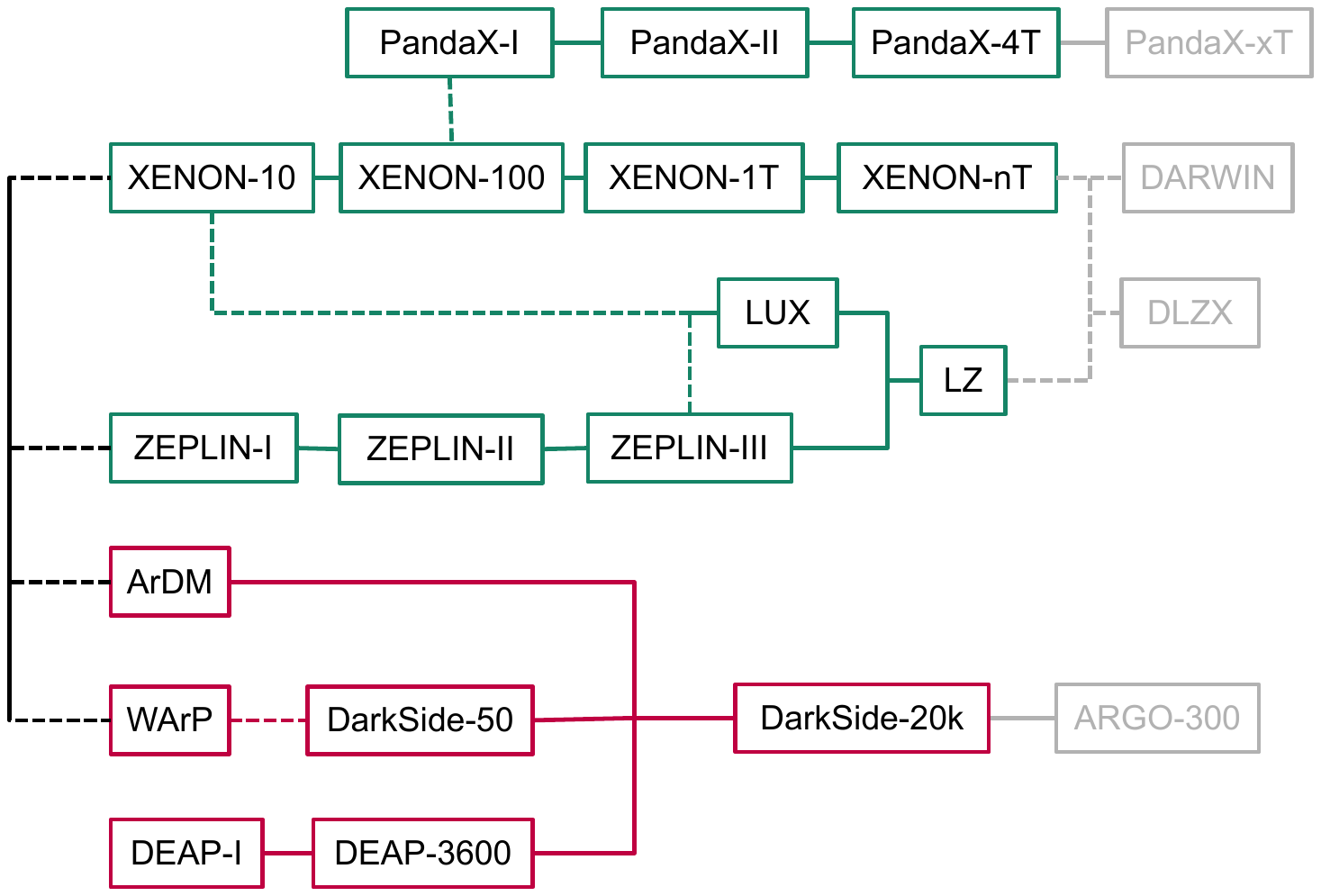}
    \caption{Genealogy of the noble gas detector. Starting from a common origin, the xenon (green) and argon (red) community split into two main branches. The dashed lines point to some weak connection between different projects, while the solid line represents the natural evolution of different detector scales. Finally, the gray diagrams represent future projects. The xenon is divided into three main projects: ZEPLIN, XENON, and PandaX. The first two will possibly evolve in a common final project. The argon community, started with three projects (ArDM, WArP and DarkSide), are currently reunited in the common project DarkSide-20k.}
    \label{fig:gen2}
\end{figure*}

\begin{figure}[ht!]
    \centering
    \includegraphics[width=\columnwidth]{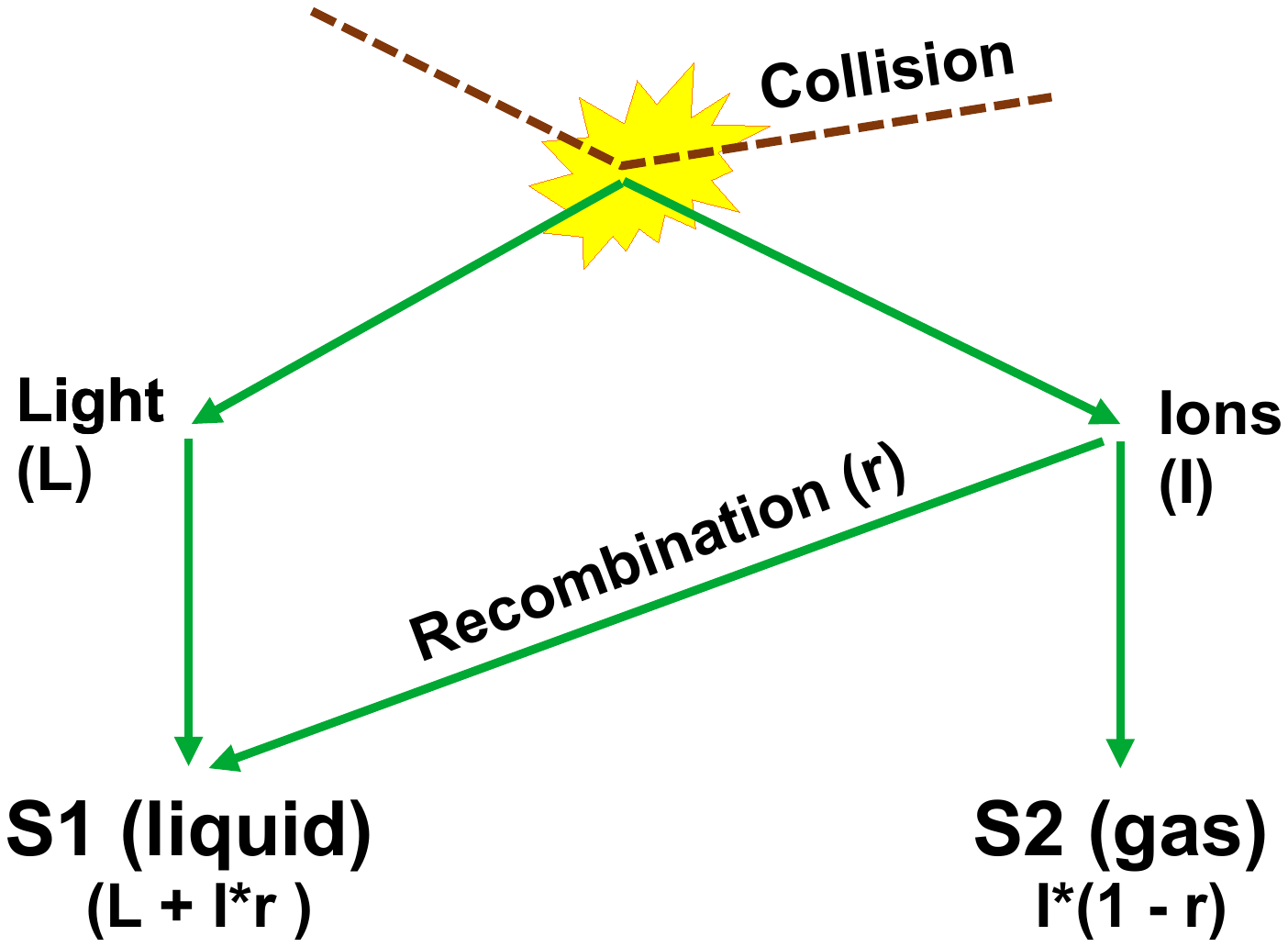}
    \caption{Scheme of the noble gas signals S1 and S2 in double-phase TCPs. A collision of an ionizing particle with the noble gas (dashed brown) produces both scintillation light and ions. In the presence of an electric field, part of the ions does not recombine producing additional light (S1), but is extracted and produces the secondary signal (S2). As a consequence, S1 and S2 are partially correlated.}
\label{fig:box}
\end{figure}

Noble gases can be operated in single phase detectors (liquid) or in double phase (liquid and gas) TPCs. In the first case, the sole scintillation signal (S1) can be used for position reconstruction and pulse shape, where possible. In the second case, for each primary interaction in the liquid target, two signals are exploited: scintillation in liquid (S1) and electro-luminescence in the top gas pocket, due to ionization electrons accelerated and extracted by strong electric fields (S2). Exploiting both S1 and S2 can help in volume fiducialization, background rejections and particle identification when S1 alone, as in the case of the xenon, is not capable of performing the ER--NR separation. A single phase detector, indeed,   is easier from a technical point of view, but in general does not allow for a high performance in event reconstruction.
S1 and S2 are usually correlated through the ion recombination process, therefore, to improve the performances, the energy estimator is defined as a linear combination of the two, whose parameters are calculated from accurate energy calibrations. A scheme of the noble gas detectors in single and double-phase is depicted in Fig.~\ref{fig:box}.

Recently, the possibility of using an ionizing only signal (S2-only) has been exploited both by argon and xenon detectors,  allowing to set a limit, after an accurate low energy background modeling, and considering the shape of Eq.~(\ref{eq:rate}) for the expected dark matter signature.  These technologies have confirmed a competitive performance also for the light dark matter detection with mass below 1 GeV~\cite{bib:dslow, bib:xelow}.

\subsection{Xenon}

The xenon has a high atomic number, and so an optimal self-shield from external background. Xenon-based detectors can search for both spin-dependent and spin-independent dark matter interactions. Even if the nuclear form factor is not that favorable~\cite{bib:lewin}, the xenon-based detectors can be a factor 5--7 smaller than argon-based detectors, mostly because of the factor $A^2$ in the cross section.

Because of the relatively higher boiling point (165 K), the xenon does not present technical issues, typical of electronic components at very low cryogenic temperatures: PMTs and the electronics chain are usually working smoothly with high performances. Even though the scintillation wavelength is in the near UV region (178 nm), the corresponding light can be easily detected by commercial photo-cathodes. It does not require wavelength shifting, resulting in high performance in event reconstruction. Since the two decay components of the scintillation light are very close to each other, S1 alone is not able to discriminate between ER and NR, while the ratio S2/S1 is usually used. Typically, the discrimination power of this classifier is one over 300. For this reason, xenon-based detectors require a high accuracy in the background control through deep purification systems and high accuracy in the material screening and selection. If not, there is a real risk of saturating the detector sensitivity in a very short exposure and creating puzzling results. Investing in the S1 only discrimination with digitizers with high sampling rate, of the order of 1 GSa/sec, could be anyway valid: even a mild preference in the NR--ER discrimination could be enough to improve the separation when combined with S2/S1 and other observables in a multivariate approach, but at present no progress has been made in this direction.

In xenon detectors it is easy to remove volatile radioactive contaminants as $^{85}$Kr, but it is more difficult to remove radon with comparable atomic mass, even if a lot of progresses has been recently accomplished in distillation techniques~\cite{bib:xent}.

\begin{figure}
    \centering
    \includegraphics[width=\columnwidth]{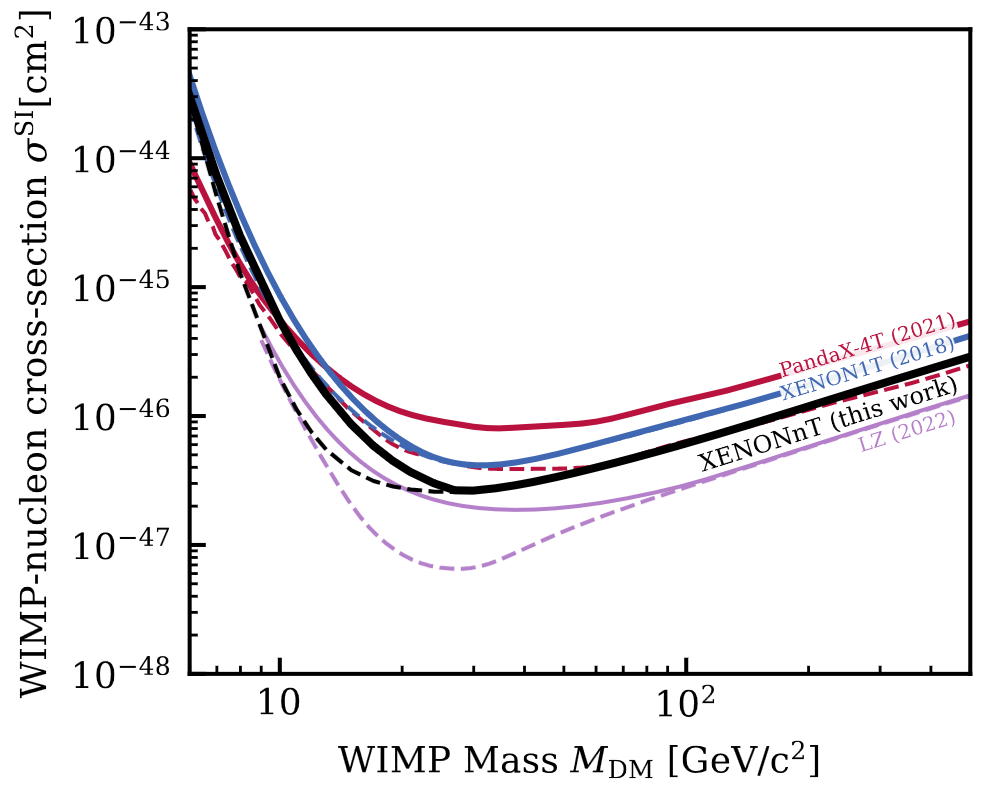}
    \caption{Comparison of the three leading xenon-based dark matter detectors, as reported in~\cite{bib:xent}, which is the publication referred to as ``this work".}
    \label{fig:xent}
\end{figure}

Figure~\ref{fig:xent} reports the most updated comparison between the leading dark matter detectors LZ, XENONnT (5.9 ton) and PandaX-4t (3.7 ton) with same order target, as reported in~\cite{bib:xent}: LZ, originated from the merging or the groups ZEPLIN and LUX is a double phase TPC, in operation since 2022 with a target of 5.5 ton~\cite{bib:lz}. XENONnT, coming from a long preceding history of versions with increasing mass (XENON10, XENON100 and XENON1T), has shown an unprecedented low background level~\cite{bib:noxexc} and is currently taking data with 5.9 ton target; PandaX-4t, is an independent collaboration that, after different preceding versions, is currently operating a 3.7 ton version~\cite{bib:panda4}.

From the experience of experiments described above, it is evident a gradual difficulty in the scaling process: a working prototype can prove the detection principle, but cannot prove the increasing technical complication coming from the increase of the target mass and the detector volume. Only a step-by-step scaling, with intermediate stages, can guarantee a solid progression of the project and a success against a highly probable failure.

The current generation of projects can reach a sensitivity very close (even though a logarithmic decade above) to the neutrino floor. All collaborations are planning future projects with targets bigger of a order of magnitude (30--50 ton) to reach a sensitivity that will basically touch the neutrino floor, as ultimate experiments on this research field. In particular, PandaX is moving independently towards a multi-ton detector, while LZ and XENON (with a possible middle scale DARWIN~\cite{bib:darwin}) are discussing a possible joint venture in a project called XLZD~\cite{bib:xlzd}.

\subsection{Argon}

The argon is lighter than xenon, therefore the self-shielding is less effective. Its boiling point is 87 K, with a corresponding technical difficulty as the one observed in the PMT electronics at this temperature~\cite{bib:aar} (even if this argument is a kind of myth: a failure of a specific PMT batch does not mean that a possible R\&D with the PMT producer would not have solved the observed issues).  The scintillation light of 128 nm is hardly matching the photo-cathode sensitivity, for this reason typically a wavelength shifter (as the TPB at 420 nm~\cite{bib:aar}) is used, with a corresponding degradation of the event reconstruction, due to the extra diffusion of light. Nevertheless, the argon shows an excellent ER--NR discrimination using S1 only: a very large separation (three order of magnitude) in the fast and slow scintillation components permits to discriminate ER from NR as one over ten billions~\cite{bib:deap}. It is worth mentioning that the argon nucleus has spin equal to zero, therefore the spin-dependent search cannot be performed, and the non-relativistic expansion of all possible relativistic operators can be done only for a reduced subset~\cite{bib:eft}.

Contrary to xenon, argon coming from the atmosphere is highly contaminated with the long-lived beta emitter $^{39}$Ar. For this reason, part of the argon community has moved towards the usage of deep underground argon in which this contaminant is reduced by a factor better than one over 1000~\cite{bib:uar}.

For the reasons described above, the argon-based detectors have a different story. Two independent lines, indeed, have emerged over the years: single phase as DEAP-3600~\cite{bib:deap} and double phase TPCs as DarkSide-50 (and marginally ArDM~\cite{bib:warp}). DEAP, operating 3.6 tons of atmospheric argon, has set the strongest limit to date for this technology, showing its intrinsic limitations.

The double phase liquid argon TPC has a more complex history.
Before DarkSide-50 with atmospheric argon (A-Ar)~\cite{bib:aar}, a smaller prototype of WArP had the only result for this technology using 2.3 liters of A-Ar~\cite{bib:warp}. This limit was further improved by the first data of underground ultra-pure argon (U-Ar) distilled from the deep Earth mantle CO$_2$~\cite{bib:uar}. After this result, DEAP-3600 currently holds the best limit for argon-based detectors (in single phase).

DarkSide-50, DEAP-3600 and ArDM are now joining a common project called GMDMC (Global argon Dark Matter Community)~\cite{bib:gadmc}. The first instance of this joint venture is DarkSide-20k, a giant TPC containing about 50 ton of U-Ar, currently under construction~\cite{bib:ds20k}. Furthermore, the GADMC is also planning a futuristic version with 300 tons of active target mass, called ARGO-300.

DarkSide-20k presents a lot of challenging aspects, and many technological novelties as compared with the existing TPCs for dark matter: first, the use of SiPM-based photo-detector modules instead of standard PMTs to overcome the issue of the cold electronics, with quite a few critical issues discussed in literature~\cite{bib:raz, bib:flash}; second, extraction and distillation of more than 50 ton of U-Ar, and preservation of its radio-purity; third, realization of a very big TPC with many challenges for high voltage, purity and event pile-up handling; fourth, use of multi-ton acrylic vessels; finally, A-Ar veto with acrylic doped with gadolinium.
Even in the case of DarkSide-20k, it is not clear why the funding agency panels have not supported intermediate scale detectors (like \emph{e.g.} 1-ton scale) with the intermediate physics goal of exploring the light dark matter mass, pushing instead for something bigger, just to fill the gap in a phantom competition with the xenon-based detectors. This choice is anyway questionable: even in case the argon technology was left behind the xenon, the unknown behavior in terms or radiopurity of multi-ton xenon-based detectors is enough worrisome to justify an alternative, even smaller but solid, with high background rejection capability, as featured by the argon.

In conclusion, as a matter of fact, the xenon-based technology has always been considered full of criticality, but in the real world it continues to be the most advanced branch of the noble gas based detectors. Liquid argon, on the contrary, is struggling to keep up and future stages are not completely clear.

\begin{table}[]
    \centering
    \begin{tabular}{|c|c|c|}
  	   \hline
  	   \textsc{Property} & \textsc{Argon} & \textsc{Xenon} \\
  	   \hline
  	   $Z$ & 18 & 54 \\
  	   $A_r$ & 39.9 & 131.9 \\
  	   $\rho$ & 1.4 g/cm$^3$ & 3 g/cm$^3$ \\
  	   $T_B$ & 87 K & 165 K \\
  	   $\lambda$ & 128 nm & 178 nm \\
  	   $\tau_{\rm fast}$ & 4 ns & 6 ns \\
  	   $\tau_{\rm slow}$ & 25 ns & 1600 ns\\
  	   LY & 40 PE/keV & 46 PE/keV \\
  	   ER--NR Classifier & S2/S2 & S1(t) \\
  	   \hline
    \end{tabular}
    \caption{Comparison between xenon and argon in dark matter detectors. From the top: atomic number, atomic weight, density, boiling point, scintillation wavelength, singlet decay, triplet decay, light yield, ER--NR classifier.}
    \label{tab:arxe}
\end{table}
\begin{figure}[h!]
    \centering
    \includegraphics[width=\columnwidth]{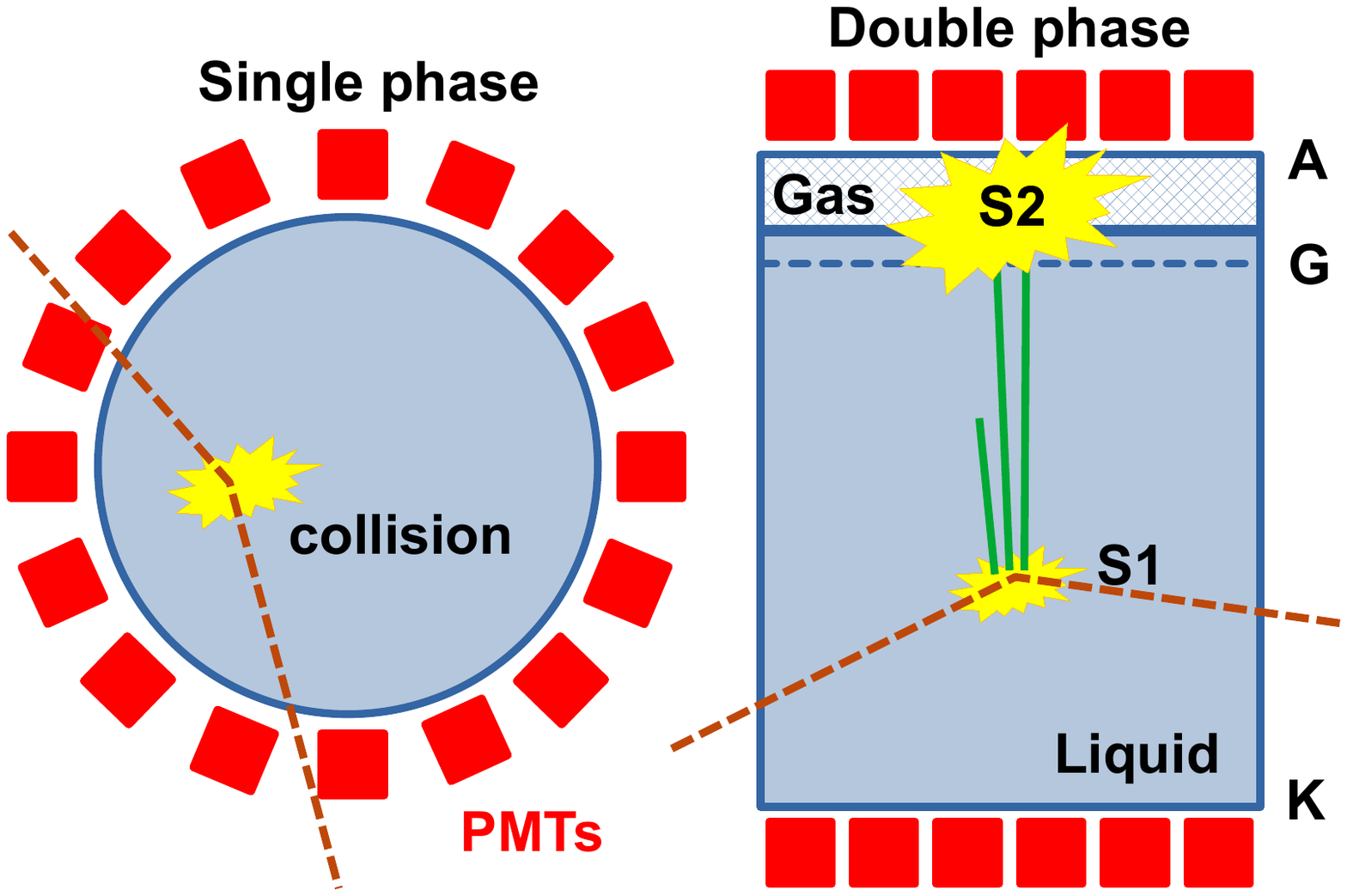}
    \caption{\emph{Left:} single phase detector in which the light emission from the scintillation process is collected by PMT's instrumented all around the detector. \emph{Right:} double phase TPC. A primary scintillation signal (S1) is produced in the liquid. The drift electric field between the cathode (K) and the grid (G) moves the ionization electrons upwards, finally  extracted by another electric field between the grid and the anode (A). The accelerated electrons in gas produce a second and stronger light signal (S2).}
    \label{fig:nobles}
\end{figure}

\subsection{Solar neutrinos}

Before moving ahead, it is worth mentioning that the multi-ton noble gas detectors have some promising by-product purposes. The very low achievable background, high target and high scintillation light yield can make them optimal detectors for precision measurement of solar neutrino fluxes coming from the proton-proton chain and the CNO cycle~\cite{bib:franco}, the two processes responsible of the hydrogen fusion in the Sun.

The current precision measurements, produced by detectors such as GALLEX/GNO~\cite{bib:gallex}, Super Kamiokande~\cite{bib:superk}, SNO~\cite{bib:sno} and especially Borexino~\cite{bib:long, bib:pp, bib:cno1, bib:cno2, bib:cno3}, have helped to better understand solar physics and neutrino oscillation. However, further improvement in precision can address plenty of other open problems, as the solar metallicity abundance~\cite{bib:cno2}, the tensions on the solar $\Delta m^2$ with reactor experiments~\cite{bib:PDG}, the precision constraint of the total solar luminosity in the low energy spectrum for extra source of energy in the Sun (as dark matter decay, indeed)  search for solar axions~\cite{bib:solax}, and non-standard neutrino interaction as smoking gun of new physics beyond SM~\cite{bib:nsi}. In other words, trying to build large dark matter detectors with some multi-purpose possibility, such as solar neutrino and also neutrino-less double beta decay detection, could in principle be reasonably acceptable.

\subsection{Others} \label{sec:other}

Besides NaI-, xenon- and argon-based detectors, there is plenty of other experiments and R\&D's using a big variety of target nuclei and techniques. Some of those, which are less sensitive to the WIMP-like particles (or sensitive only to light masses) are not discussed for the purpose of present articles. It is interesting to mention some of them, which are playing an important role in direct dark matter search: CRESST using a target with CaWO$_4$~\cite{bib:cresst}, CDMSLite and  SuperCDMS~\cite{bib:cdmsS} using germanium, DAMIC using silicon~\cite{bib:damic},  PICO--60 using C$_3$F$_8$~\cite{bib:pico60}, and NEWS-G using neon~\cite{bib:newsg}.

If one may think that DAMA, made of sodium  (light nucleus, $A \simeq 23$) and iodine (heavy nucleus, $A \simeq 127$) cannot be compared to xenon (heavy nucleus, $A \simeq 131$) because the origin of the annual modulation comes from the interaction with sodium rather than iodine, one should also think to what happens with a big variety of other atoms, used by other experiments, with null results. And they are made of carbon ($A \simeq 12$), oxygen  ($A \simeq 16$), fluorine  ($A \simeq 19$), calcium  ($A \simeq 40$), argon  ($A \simeq 40$), tungsten  ($A \simeq 183$), germanium  ($A \simeq 73$), silicon  ($A \simeq 28$) and neon ($A \simeq 20$). Results do not easily reconcile even in case of spin-dependent (SD) interactions, as \emph{e.g.} the xenon contains approximately the same percentages of isotopes with nuclear spin 0, 1/2 and 3/2.
Considering the full list of nuclei tested so far in the framework of SI (and also SD) interaction with SHM, it is extremely hard to believe that sodium plays such a special role among all the elements in the periodic table.

\section{Evolution of results} \label{sec:evol}

A good way to explore the evolution of results on direct detection of dark matter can be done reading the dark matter review as reported by the Particle Data Group (PDG)~\cite{bib:pdgweb}. Downloading old versions from 1996 to the latest update (2022) (see \cite{bib:p1996, bib:p1998, bib:p2000, bib:p2002, bib:p2004, bib:p2006, bib:p2008, bib:p2010, bib:p2012, bib:p2014, bib:p2016, bib:p2018, bib:p2020, bib:PDG} and Refs. therein ), one can see that the Review has increasingly dedicated a larger and larger number of pages, passing from about 4 to more than 30 in about 25 years. This fact, of course, is not only related to the increasing interest in dark matter, but also to the overall increase of the space dedicated to physics reviews in PDG. A correct comparison should be somehow normalized.

The first SI $\sigma$--$M_\chi$ plot appeared only in 2010 and  it was updated on 2011~\cite{bib:p2010}. In this paradigm, one can immediately see the strong tension of the two DAMA solutions (as interpreted in the SI framework by~\cite{bib:damaSI}), made of two islands (one for the sodium solution and one for the iodine solution), with the other experiments, as XENON-100, EDELWEISS and CMDS-Si. Along with DAMA, another small-scale detector based on germanium and called CoGeNT~\cite{bib:cogent} shows a similar solution close to the DAMA sodium regions. In the same plots, it appears the region in which possible SUSY candidates can be potentially discovered, as hidden channels, at LHC.

\begin{figure}[!ht]
    \includegraphics[width=\columnwidth]{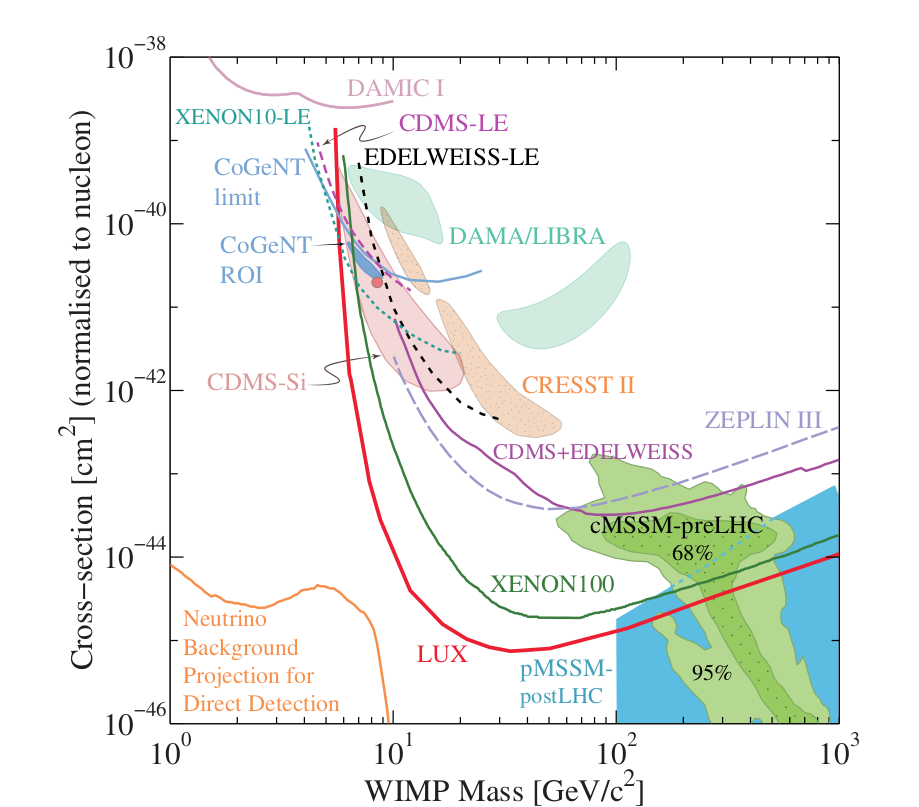}
    \caption{Cross section as a function of the WIMP-like mass in the SI framework, as reported in PDG 2013. In the same plot, limits (solid curves) and positive results (islands with a given CL) are reported for various experiments, see text.}
    \label{fig:PDG13}
\end{figure}

In 2012 nothing was really changed, but in the 2013 update~\cite{bib:p2012} (see Fig.~\ref{fig:PDG13}), persisting also in 2014~\cite{bib:p2014}, new islands appeared, very close to DAMA: both CDMS-Si and CRESST had an excess over the predicted background. All these positive results appeared singularly close to the Higgs boson discovery in July 2012 at LHC~\cite{bib:higgs1, bib:higgs2}. At the same time, the absence of evidence of SUSY candidates started to be digested by the scientific community, but still non widely consolidated~\cite{bib:mssm}.

\begin{figure}[!ht]
    \includegraphics[width=\columnwidth]{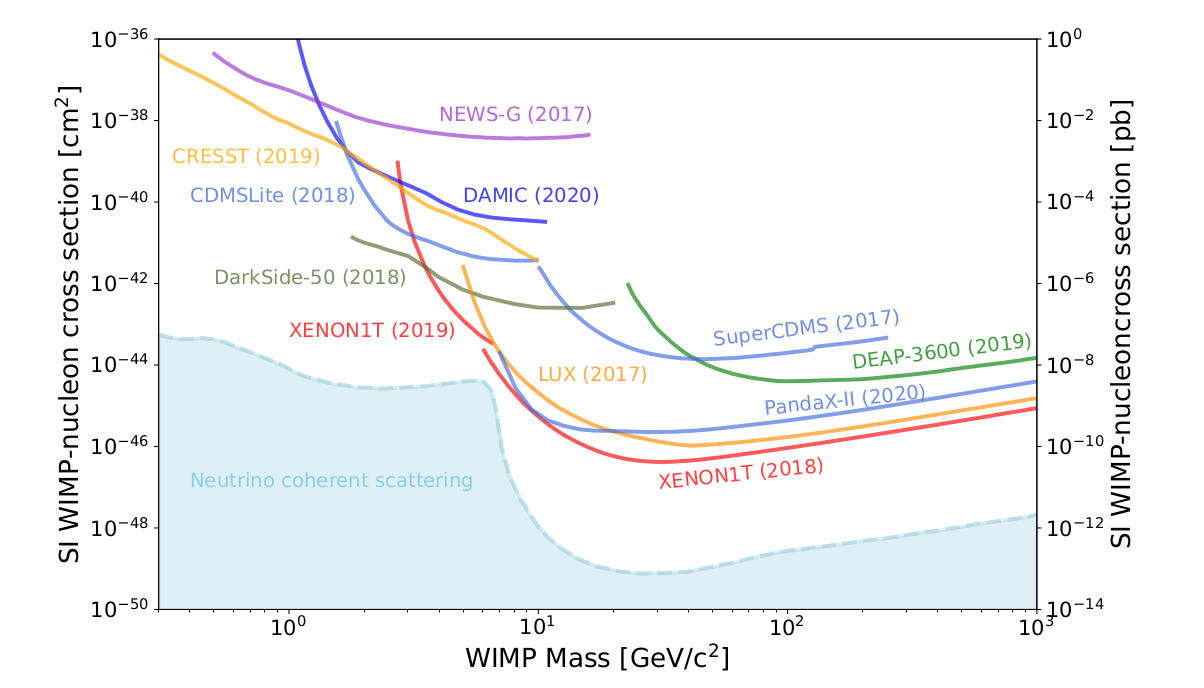}
    \caption{Cross section as a function of the WIMP-like mass in the SI framework, as reported in PDG 2022. Only limits (solid curves) are reported by various experiments, see text.}
    \label{fig:PDG22}
\end{figure}

In 2015, CRESST-II observed no excess, and then the previous islands due to CaWO$_4$ were removed in the updated PDG~\cite{bib:p2014}. In 2016 nothing was really changing, except for some improvement on previous limits by many experiments~\cite{bib:p2016}. In the 2017 update~\cite{bib:p2016}, from null results by further upgraded versions of CoGeNT, the corresponding island was removed, and new limits, such as PICO-60 and DEAP-3600 were added. In the 2018 update~\cite{bib:p2018}, coinciding with important  releases from ANAIS-112 and COSINE-100, the DAMA islands were removed from the PDG. Notice that the DAMA SI islands are almost never presented in the official DAMA publications, and recently survive mainly in the publications of the NaI-based experiments trying to reproduce DAMA.
Hereafter, up to the last upgrade in 2022~\cite{bib:PDG}, the PDG SI $\sigma$--$M_\chi$ plot reports only upgraded limits (see Fig.~\ref{fig:PDG22}), closer and closer to the neutrino floor, for both low and high masses.

Recently, since the statistical fluctuations of the neutrino floor can become important in present and next generation of dark matter experiments, the name has been changed to \emph{neutrino fog} or \emph{mist}, considering the real impact of how this expected background grows with the experimental exposure~\cite{bib:mist}.

The common practice of focusing on the SI $\sigma$--$M_\chi$ plot has received some criticisms. One may think that comparing all experimental results in the same SI $\sigma$--$M_\chi$ plot could not be a comprehensive and accurate way to address the dark matter problem, and would be only a generally subjective, limited and imprecise action. One can reasonably accept this criticism, but it remains still unexplained how a unique positive result (from DAMA) can be compatible, independently of the model,  with a tens of other null results made by experiments of comparable or larger sensitivity and using even similar nuclei as discussed above. Those experiments are not detecting any positive signal anyway, regardless of the fact that nature has chosen a SI interaction, or whatever, for visible and dark matter particles.  

As has happened many times in the history of physics, the research is going through a hard period in which the knowledge is stuck and experiments are becoming more and more challenging and expensive. The whole scientific community is moving by inertia, after the thrust of a strong theory that is now left behind the shoulders, and becomes everyday fuzzier and fainter. Moreover,  this motion proceeds so smoothly that someone could not even realize how he ended up in it.

Furthermore, the belief is so strong that people are already thinking about the future made, not only of bigger and bigger detectors capable of reaching the neutrino floor, but also thinking about how to drive into the neutrino fog with a huge detector capable of exploiting the directionality of the dark matter. The latter in particular looks very futuristic, given the present status of results.  

From recent history, it is clear that there was some original enthusiasm, supported by a very appealing theory as SUSY, and close to Higgs discovery there was a cluster of experimental ``excesses'', leaded by the DAMA result that has probably amplified the expectation that dark matter particles could be really detectable.

It is anyway commonly accepted that reaching at least the neutrino floor is a kind of ``moral duty'' for the scientific community, before trashing completely WIMPs and all WIMP-like paradigms all at once, in favor of other particle or non-particle solutions of the problem of the missing mass of the Universe.
\begin{figure}[!ht]
    \includegraphics[width=\columnwidth]{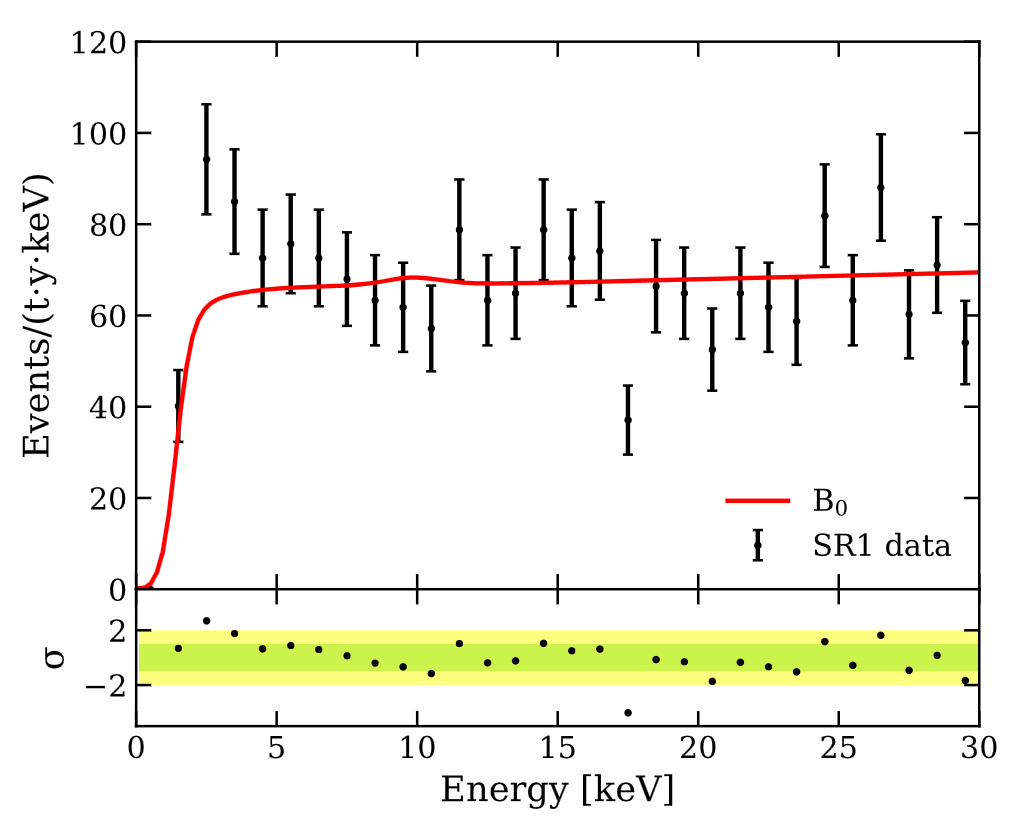}
    \caption{Excess of events above the threshold reported by XENON1T, compared with their background model (red).}
    \label{fig:xecc}
\end{figure}

Finally, to conclude with another interesting sociological case, one can remind when XENON1T published a presumed excess above the threshold as reported in Fig~\ref{fig:xecc}~\cite{bib:xexc}, and later non confirmed by XENONnT with a lower overall background~\cite{bib:noxexc}. The collaboration was also warning that such a spectrum could even come from an unaccounted tritium beta spectrum.
Regardless of whether or not some trained eyes see these two outlier points above the event threshold as an excess, it is curious that these two points have received about 600 citations to date, mostly from theorists and phenomenologists.

What can one do as a particle physicist? The choice of spending some more time in this kind of search will for sure pay back, as advances in knowledge and technological implications are granted by basic research. Anyway, one should not exaggerate, because if the main purpose of an experiment are its secondary goals, in this case one can propose to search for whatever non-falsifiable theory, and this would make the science become a practical paradox.

\subsection{A Drake equation}

The Drake equation is a probabilistic argument used to estimate the number of active, communicative extraterrestrial civilizations in the Milky Way galaxy~\cite{bib:drake}. Of course, this equation is more useful for ``understanding" rather than ``quantifying", as its parameters are affected by large uncertainty and supported sometimes by naive arguments.

One can imagine to follow the same approach for the probability $\mathcal{P}_\chi$ that dark matter is made of WIMP-like (or light WIMP) particles and can be detected by experiments on the Earth.

A possible ansatz, containing the main terms, is:
\begin{equation}
    \mathcal{P}_\chi = f_{\rm e}\cdot f_{\rm s}\cdot f_{\rm th} \cdot f_{\rm exp} \cdot f_{\rm det} \cdot f_{\Omega}
\end{equation}
where:

$f_{\rm e} =$ fraction of energy in the full mass interval for dark matter candidates. As there is no actual reason to prefer one candidate to another, one can estimate this ratio as 5/89, where 5 corresponds in the logarithmic scale to 0.1--10000 GeV/c$^2$ interval over the full range, from axions to massive primordial black holes.

$f_{\rm s} =$ fraction of possible cross section available for a WIMP-like interaction. If one considers the full range, from the already reached ($\log(\sigma)=-46$) down to the squared Plank mass ($\log(\sigma)=-66$) of 20 orders of magnitude, and that this kind of candidates cannot be much lower than the neutrino floor ($\log(\sigma)=-50$) for experimental reasons, this factor has to be taken as 4/20.

$f_{\rm th}=$ probability that the missing mass of the Universe is explained by particles or by some acceleration anomaly, \emph{i.e.} the failure in extrapolating the Newton law (General Relativity) for the Solar System ($10^8$ km) to the galactic scale (kpc), or something else. There is no real reason why this fraction should not be at least 1/2, or even lower (1/3).

$f_{\rm exp}=$ probability that only one experiment (DAMA) has detected dark matter over about 10 with comparable sensitivity, that is about 1/10.

$f_{\rm det}=$  probability that the dark sector cannot exchange information with the visible sector with other (weak) interaction but gravity, as in some versions of the so-called Mirror Matter models~\cite{bib:mirror}. This probability can be set to 1/2, as there is not a real reason why the dark matter should share the same interaction properties of the visible matter.

$f_{\Omega}=$ catastrophic probability that the Big Bang cosmology is all wrong. Nobody will ever admit that, but also saying that the Big Bang cosmology is 100\% correct,  would look weird as well. Besides static or quasi-static physical cosmological models and the like, one could consider alternative and radical views of the Universe, such as the Simulation hypothesis~\cite{bib:simul} or the Mathematical Universe hypothesis~\cite{bib:max}. In this case, one would reasonably accept a 50\% chance from a philosophical point of view, even though one may also think that dark matter  could anyway be included in this kind of Simulation. Furthermore, this probability could be correlated to $f_{\rm th}$, if modification of gravity can explain the missing mass. Anyway, assuming one for this factor is part of the game, even if quite disturbing.  

With those basic terms, one gets about 1:3560 for $\mathcal{P}_\chi$. If the equation is empowered with other terms, each of them will be $\lesssim 1$, then the updated probability will be likely  less than this first estimation. Now it is time to bet.

\subsection{Mala tempora}

Until the early 80's of the past century, new particles came out of accelerators every day, and it was relatively easy to understand the hidden logic among particles and interactions in the frameworks of quantum field theory. Theorists got carried away and fantasized about many extensions of SM, which, although constituting an excellent description of the observed phenomena, left and still leaves indications of a more complete high-energy theory. And SUSY, with its by-product ``WIMP miracle''~\cite{bib:miracle}, was the most awaited guest at the party.

LHC, close to the discovery of the Higgs boson, has shown not to be really suitable as a discovery machine for the new physics beyond SM. It is therefore probable that the planned high luminosity stage~\cite{bib:hllhc} will end up in controversial \emph{anomalies} and \emph{tensions}, which will only complete an already  existing long list. It would have been probably better to shut down LHC and speed up the construction of the Future Circular Collider, also known as FCC, with a center-of-mass collision energy of 100 TeV, that is almost one order of magnitude higher than LHC. To be honest, one should also admit that, in principle, there is no indication that a possible new physics emerges just at 100 TeV rather than at 1 PeV or more.
Therefore, it is anyway like sailing in the open sea without knowing if and where the next land will be. If one can speak of a possible ``crisis in modern particle physics", now is precisely that moment. The only (weak) hope, in light of the phrase ``mala tempora currunt", is the memory that in similar situations in the history of physics, a significant revolution often followed a deep crisis.

The only serious risk is that, in the absence of concrete scientific objectives, experimental collaborations may become uncontrollable inefficient organizations, whose main goals can be something different from scientific research.

\subsection{Falsifiable and Scientific}

Karl Popper in his book \emph{The Logic of Scientific Discovery} (1934) suggested that a statement, a hypothesis, or theory, to be considered scientific, should be ``falsifiable'', \emph{i.e.} logically contradicted by an empirical test. The material implication  ``{\bf IF} it is scientific {\bf THEN} is falsifiable", \emph{i.e.},
\begin{equation}
  \textsc{scientific} \rightarrow \textsc{falsifiable},  
\end{equation}
leaves a wide room to theories that are falsifiable but not scientific. Proposing a massive particle detectable in underground experiments is for sure falsifiable, but not necessarily scientific, especially if there is no theory behind, and no clear motivation why it should be worth searching for it. Reading the material implication in the opposite direction (as basic logical fallacy) has sometimes made a lot of confusion, not only for the dark matter case, but also for numerous extensions of SM, based on aesthetic argument  instead of real necessity. This misunderstanding becomes even more threatening when the properties of a given dark matter candidate are updated after the initial dark matter candidate is not found in the place in which it was proposed~\cite{bib:sabine}.

This is the case of WIMP particles emerging from SUSY. The Minimal Supersymmetric Standard Model (MSSM), introduced to accommodate the problem of the hierarchy of the Higgs mass, should have broken at the Higgs mass scale ($\sim 100$ GeV) and should have predicted  the existence of stable massive dark matter candidates. The absence of a SUSY particles discovery at the LHC has pushed theorists to abandon the ``naturalness'' concerns  ~\cite{bib:susyhigh}, add other parameters and mechanisms, and increase the SUSY breaking scale, creating a big family of X-MSSM models, where X stands for the acronym of the case, see~\cite{bib:xmssm} and Refs. therein. The PDG has recently removed those families of allowed regions in SI dark matter parameter space~\cite{bib:PDG}, being already halved by xenon-based dark matter detectors. Furthermore, saying that there is still a 50\%, or so, unexplored region is, for what is discussed, definitely pointless.

What one needs is a change of paradigm, and this is very well summarized by F. Nesti \emph{et al.}~\cite{bib:nesti}:
\begin{quote}
``In detail, we advocate for a paradigm according to which, after abandoning the failing $\Uplambda$CDM scenario, we must be poised to search for scenarios without requiring that: (a) they naturally come from (known) “first principles'' (b) they obey to the Occam razor idea (c) they have the bonus to lead us towards the solution of presently open big issues of fundamental Physics. On the other side, the proper search shall: (i) give precedence to observations and the experiment results wherever they may lead (ii) consider the possibility
that the Physics behind the Dark Matter phenomenon be disconnected from the Physics we know and and does not comply with the usual canons of beauty''.
\end{quote}

This strategy is quite reasonable. However, the discovery of some elusive particle solution in underground laboratories, not supported by any theoretical framework, will be a big deal from an epistemological point of view, but anyway better than nothing.

\subsection{The emperor is naked!}

The dark matter in shape of WIMP or WIMP-like or light WIMP particles is
 widely accepted as true or professed as being plausible, due to an unwillingness of the general scientific community to criticize it or be seen as going against the mainstream opinion.
 
 The \emph{Emperor's New Clothes} fairy tale by Hans Christian Andersen is a metaphor about logical fallacies, that reads in this case: no one believes in such a dark matter, but everyone believes that everyone else believes in it, until some child comes out of the crowd and shouts: ``the emperor is naked"! But, in this case, such a child, if ever any, has yet to be born.  

\subsection{A way out}

Even though direct searches have not solved the dark matter problem, there are a few interesting things to do. First, one might need to rethink our understanding of gravity. MOND, or similar, is one idea, essentially saying that our current gravitational laws might not apply on larger scales. Alternatively, one might discover new particles that interact even more weakly than previously thought, making them incredibly hard to detect.

Another possibility is that one is missing something in our theoretical framework. Maybe our understanding of particle physics needs an upgrade. Some theories propose new types of particles, like sterile neutrinos or axions, which could be potential dark matter candidates. These particles would be elusive and might interact with normal matter in ways that are not fully understood yet. An interesting change of paradigm is proposed here~\cite{bib:mrad}.

On the observational side, upcoming experiments, like the James Webb Space Telescope~\cite{bib:jwst} and Euclid~\cite{bib:euclid} might provide more insights into the cosmic web and the distribution of matter in the Universe. High-energy cosmic-ray observatories and gravitational wave detectors could also bring new clues.

In essence, solving the dark matter puzzle might require a combination of refining our theories, developing more sensitive detection methods, and pushing the boundaries of our observational capabilities.

\section*{Conclusions}

We do not want to spoil the party, so we will conclude by reiterating the conviction that the current direct detection of dark matter in the region of the WIMP-like and light dark matter is absolutely valid, as the cost/benefit is still affordable and justifiable. About the next generation of detectors, the same cannot be said with the same certainty.

However, we want to point out that this quest is currently moved by inertia after the strong thrust impressed a few years ago by a solid theoretical framework that now is
becoming farther and fainter. It happened many times in the history of physics that some puzzle has been solved because there was a clear indication where to search for its solution. Now it looks as though it is not the case for dark matter any more.

The missing mass of the Universe, explained as a heavy particle forming a halo all around galaxies is an simple theory, so simple and easy to attract many scientists who may be  not so willing to dwell in complicated calculations and ideas. But it is also persisting, because we are led to think that easier solutions are always the ones chosen by nature.

The fact that the presence of the missing mass of the Universe is supported by so many irrefutable pieces of evidence makes the direct search of dark matter a kind of moral duty that we seem to pursue at all costs. But this idea is pointless, and probably dangerous.

In this report, we have seen that the leading role in the dark matter search has been driven first by the DAMA annual modulation result, followed by other results singularly close to the discovery of the Higgs boson. This apparent convergence slowly disappeared in the subsequent years, and was overcome by the liquid noble gas detectors, which are continuing showing null results, and increasing their sensitivity with bigger and bigger targets.

Given the very small chance of detecting such dark matter particle candidates, as inferred by a procedure similar to the Drake equation, we can conclude that the present stage of the quest, without getting sidetracked by science fiction-like projects, is anyway necessary to start rethinking the Universe.

\section*{Acknowledgements}

First of all, we would like to thank V. Fano and his team from University of Urbino: the origin of this work can somehow be traced back to the interaction with him.
A lot of pieces of information for Fig.~\ref{fig:gen} and~\ref{fig:gen2} are coming from colleagues, as they are not officially reported in collaboration articles.
We would like to thank in particular, G. Di Carlo, L. Grandi, A. Ianni, N. Di Marco, K. Pelczar and C. Vignoli for useful discussions about the dark matter problem and for helping with the reconstruction of the experiment genealogies.
Finally, we would like to thank G. Ranucci, F. Nesti, A. Di Giovanni and R. Biondi, for comments, suggestions and proofreading.

\vspace{0.5cm}
\noindent
\rule{\columnwidth}{0.4pt}

{\footnotesize
\section*{List of acronyms}

\noindent
LHC = Large Hadronic Collider

\noindent
$\Uplambda$CMD = Lambda Cold Dark Matter

\noindent
MOND = Modified Newtonian Dynmics

\noindent
SUSY = Super Symmetry

\noindent
WIMP = Weakly Interacting Massive Particle

\noindent
GUT = Grand Unification Theory

\noindent
MACHO = Massive Astrophysical Compact Halo Object

\noindent
ALP = Axion-Like Particle

\noindent
SI = Spin Independent

\noindent
SD = Spin Dependent

\noindent
SM = Standard Model

\noindent
SHM = Standard Halo Model

\noindent
CL = Confidence Level

\noindent
PMT = Photo-Multiplier Tube

\noindent
ER = Electron Recoil

\noindent
NR = Nuclear Recoil

\noindent
R\&D = Research and Development

\noindent
SiPM = Silicon Photo-Multiplier

\noindent
TPC = Time Projection Chamber

\noindent
S1 = Primary scintillation light in TPCs

\noindent
S2 = Secondary scintillation light in TPCs

\noindent
UV = Ultra Violet

\noindent
FCC = Future Circular Collider

\noindent
MSSM = Minimal Supersymmetric Standard Model
}

% References

\end{document}